\documentclass[aps,prd,nofootinbib,showpacs,superscriptaddress]{revtex4} 
\usepackage[sumlimits]{amsmath}
\usepackage{epsfig}
\usepackage{amssymb}
\usepackage{graphicx}
\usepackage{pstricks}
\usepackage{color}
\usepackage{bbold}
\usepackage{hyperref}
\usepackage{ragged2e}

\newsavebox{\measurebox}

\begin{document}

\title{Dynamical and observational constraints on the Warm Little Inflaton scenario}
\author{Mar Bastero-Gil}
\email{mbg@ugr.es}
\affiliation{Departamento de F\'{i}sica T\'{e}orica y del Cosmos, Universidad de Granada, Granada-18071, Spain}
\author{Arjun Berera}
\email{ab@ph.ed.ac.uk}
\author{Rafael Hern\'{a}ndez-Jim\'{e}nez}
\email{s1367850@sms.ed.ac.uk}
\affiliation{School of Physics and Astronomy, University of Edinburgh, Edinburgh, EH9 3FD, United Kingdom}
\author{Jo\~{a}o G. Rosa}
\email{joao.rosa@ua.pt}
\affiliation{Departamento de F\'{i}sica da Universidade de Aveiro and CIDMA, Campus de Santiago, 3810-183 Aveiro, Portugal}

\begin{abstract}
We explore the dynamics and observational predictions of the Warm Little Inflaton scenario, presently the simplest realization of warm inflation within a concrete quantum field theory construction. We consider three distinct types of scalar potentials for the inflaton, namely chaotic inflation with a quartic monomial potential, a Higgs-like symmetry breaking potential and a non-renormalizable plateau-like potential. In each case, we determine the parametric regimes in which the dynamical evolution is consistent for 50-60 e-folds of inflation, taking into account thermal corrections to the scalar potential and requiring, in particular, that the two fermions coupled directly to the inflaton remain relativistic and close to thermal equilibrium throughout the slow-roll regime and that the temperature is always below the underlying gauge symmetry breaking scale. We then compute the properties of the primordial spectrum of scalar curvature perturbations and the tensor-to-scalar ratio in the allowed parametric regions and compare them with Planck data, showing that this scenario is theoretically and observationally successful for a broad range of parameter values.
\end{abstract}
 \maketitle


\section{Introduction}

The inflation paradigm \cite{inflation} remains as the most appealing mechanism to explain the present flatness, homogeneity and isotropy of the observable universe. In addition, fluctuations generated during an early phase of inflation yield a primordial spectrum of density perturbations, which is nearly scale invariant, adiabatic and gaussian, in agreement with cosmological observations. In the standard picture of ``cold inflation" (CI), the state of the universe is the vacuum state, since accelerated expansion quickly erases all traces of any pre-inflationary matter or radiation density. This leads, however, to a supercooled universe and the need to explain the transition from inflation to the ``hot Big Bang" state required  by Big Bang Nucleosynthesis and the physics of recombination leading to the Cosmic Microwave Background (CMB) that we observe today. This transition necessarily requires the (partial) decay of the inflaton into ordinary matter and radiation and thus to its interactions with other fields. 

Such interactions are conventionally thought to play a negligible role during the slow-roll phase of inflationary models driven by a scalar inflaton field. The reasoning behind this na\"ive expectation lies in the fact that the perturbative decay width of particle is generically smaller than its mass, which in turn lies below the Hubble expansion rate for a slowly rolling scalar field. Hence, in this picture particle production can never compete with the inflationary expansion rate and inflaton decay can only play a significant role at the end of the slow-roll regime, leading to the standard ``reheating" paradigm. One must note, however, that the perturbative decay width only describes the decay of a field close to the minimum of its potential \cite{Graham:2008vu}, which is obviously not the case during slow-roll inflation, and that finite temperature effects can further significantly enhance the rate at which the inflaton dissipates its energy into other degrees of freedom. Taking this into account, one can conceive an alternative {\it warm inflation} (WI) paradigm \cite{Berera:1995wh,Berera:1995ie}, where dissipative effects and associated particle production can, in fact, sustain a thermal bath concurrently with the accelerated expansion of the Universe during inflation.

Interactions with an ambient thermal bath generically lead to non-equilibrium effects in the dynamics of a scalar field. For a field evolving slowly compared to the characteristic time scale of the thermal bath, the leading non-equilibrium effect is a dissipative friction term $\Upsilon\dot\phi$ in its equation of motion \cite{Berera:2008ar}, where $\Upsilon=\Upsilon(\phi,T)$ can be computed from first principles given the form of the interactions between the scalar field and the thermalized degrees of freedom. One can easily show that, for a homogeneous field, this implies the continuity equation $\dot\rho_\phi + 3H(\rho_\phi+p_\phi)=-\Upsilon\dot\phi^2$, such that overall energy-momentum conservation implies the existence of an identical term with opposite sign in the continuity equation for the thermal fluid. One can show explicitly that dissipative effects in the inflaton's equation of motion lead to particle production in the thermal bath, which prevents the exponential dilution of the latter in a quasi-de Sitter background.  In particular, for a nearly-thermal relativistic fluid, i.e.~radiation, we have:
\begin{equation}
\dot{\rho}_{R}+4H\rho_{R} = \Upsilon\dot{\phi}^{2}~,
\end{equation}
such that for the slowly evolving inflaton field, the dissipative source term 
on the right hand side remains nearly constant throughout inflation, yielding a slowly evolving radiation energy density $\rho_R= C_R T^4\simeq \Upsilon\dot\phi^2/4H$, where $C_R=\pi^2 g_* /30$ for $g_*$ relativistic degrees of freedom in the thermal bath. The resulting nearly constant temperature during inflation exceeds the de Sitter Hawking temperature $\sim H$ for 
\begin{equation}
{\dot\phi^2/2\over V(\phi)} > {2C_R\over 9} Q^{-1} {H^2\over M_P^2}~,
\end{equation}
where $V(\phi)$ denotes the scalar potential and $Q=\Upsilon/3H$. Since $H\ll M_P$ in most inflationary models, one can obtain a warm inflationary universe, $T\gtrsim H$, consistently with a slow-roll evolution, even for weak dissipative effects, $Q\ll 1$. One can further show that the radiation energy density can never exceed the inflationary potential in a slow-roll regime, guaranteeing a period accelerated expansion:
\begin{equation} \label{radiation_abundance}
{\rho_R\over V(\phi)} \simeq {1\over 2} {\epsilon_\phi\over 1+Q} {Q\over 1+Q}~,
\end{equation}
where we have used the conventional slow-roll parameter 
$\epsilon_\phi = M_P^2(V'(\phi)/V(\phi))^2/2$, such that consistence of 
the slow-roll evolution requires $\epsilon_\phi <1+Q$. This in turn also 
implies that, at the end of the slow-roll regime, when 
$\epsilon_\phi \sim 1+Q$, one may attain $\rho_R\sim V(\phi)$ if a 
strong dissipative regime $Q\gtrsim 1$ can be achieved. In such cases 
radiation will smoothly become the dominant component at the end of 
inflation, providing the necessary ``graceful exit" into 
the ``hot Big Bang" cosmic evolution \cite{Berera:1996fm}. Although 
there may be additional particle production at the of inflation, no 
reheating is actually necessary in warm inflation under the above conditions.

In addition to the natural exit from inflation, warm inflation exhibits 
several attractive features that have been explored in recent years. 
For instance, the dissipative friction damps the inflaton's 
evolution, making slow-roll easier or, equivalently, alleviating the conditions on the flatness of the inflaton potential, expressed now by the slow-roll conditions $\epsilon_\phi, |\eta_\phi|\ll 1+Q$, where $\eta_\phi=M_P^2 V''(\phi)/V(\phi)$. This may potentially provide a solution to the so-called ``eta-problem" typically found in string/supergravity inflationary models where generically $\eta_\phi \sim \mathcal{O}(1)$ \cite{Berera:1999ws,BasteroGil:2009ec}. Analogously, this may also alleviate the need for large (superplanckian) inflaton field values in chaotic inflation models.  There are various other features of warm
inflation that have been explored in the literature 
\cite{Graef:2018ulg, Gron:2018rtj,Oyvind Gron:2016zhz,Rangarajan:2018tte,Li:2018wno,Herrera:2018cgi,Arya:2017zlb,Herrera:2017qux,Videla:2016ypa,Peng:2016yvb,Goodarzi:2016iht,Levy:2016jfh,Visinelli:2014qla,Herrera:2014mca,Setare:2013dd,Bastero-Gil:2013owa, Cerezo:2012ub,BasteroGil:2011cx}.

The fluctuation-dissipation theorem is behind one of the most attractive
features of warm inflation, since it implies a noise term in the
inflaton equation that directly sources fluctuations in the inflaton field
and consequently modifies the resulting primordial spectrum of curvature
perturbations that later becomes imprinted in the
CMB \cite{Berera:1995wh,Berera:1999ws, Hall:2003zp, Moss:2007cv,
Graham:2009bf, Ramos:2013nsa, Bastero-Gil:2014jsa}.
This generically enhances the amplitude of scalar curvature perturbations
while leaving tensor modes unaffected, due to the weakness of gravitational
interactions with the thermal bath, therefore lowering the tensor-to-scalar
ratio with respect to ``cold inflation" scenarios.
This feature is intrinsic to warm inflation models and it was
shown explicitly in \cite{BasteroGil:2009ec}, well before the BICEP and
Planck
results, that the presence of radiation and dissipation suppresses
the tensor-to-scalar ratio.  That paper \cite{BasteroGil:2009ec}
computed the tensor-to-scalar ratio in the monomial
$\phi^2$ and $\phi^4$ models
and was the only analysis at the time that predicted for these models a low
tensor-to-scalar ratio, which now we see is consistent with data.
Subsequent work
further developed the analysis
\cite{Cai:2010wt, Bartrum:2013fia}. Inflaton fluctuations may also be in a
thermal rather than vacuum state as a consequence of the coupling to the
radiation bath and, depending on the form of the dissipation coefficient,
the interplay between inflaton and radiation perturbations may also lead to
growing modes in the spectrum \cite{Moss:2007cv}. Overall, this means that
CMB observations can be used to probe the interactions between the inflaton
and other fields, which is not possible in conventional models with a
separate reheating period. Warm inflation thus provides a new arena to probe
high energy fundamental physics.

Realizing warm inflation within a consistent quantum field theory framework has, however, proved to be a challenging endeavor. Non-equilibrium dissipative effects are Boltzmann suppressed unless the particles in the radiation bath are relativistic, while the inflaton typically gives a large mass to the fields it couples directly to. In addition, relativistic particles change the form of the inflaton potential at finite temperature, typically inducing large thermal corrections to the inflaton's mass that may prevent slow-roll unless the associated inflaton couplings are very suppressed, therefore rendering dissipative effects ineffective in sustaining a thermal bath during inflation \cite{BGR, YL}. For several years, the leading solution to these problems was to consider models where the inflaton only couples directly to heavy fields, which in turn decay into light particles in the thermal bath \cite{Berera:2002sp}. In these scenarios thermal corrections to the inflaton potential become Boltzmann-suppressed, while dissipative effects can nevertheless be significant if one considers a large number of fields coupled to the inflaton\footnote{In this case dissipative effects are the result of heavy virtual modes that are not Boltzmann-suppressed \cite{Moss:2006gt, BasteroGil:2010pb, BasteroGil:2012cm}.}. While such scenarios may find natural realizations in specific constructions within e.g.~string theory \cite{BasteroGil:2011mr} where field multiplicities can be large during inflation, they cannot provide a simple and sufficiently generic realization of warm inflation.

A more promising scenario was proposed recently \cite{Bastero-Gil:2016qru} where the above-mentioned problems were addressed using symmetries rather than large field multiplicities. This {\it Warm Little Inflaton} (WLI) scenario, so-called due to its similarities with Little Higgs models of electroweak symmetry breaking \cite{ArkaniHamed:2001nc, Schmaltz:2005ky}, considers an inflaton field that corresponds to the relative phase between two complex Higgs scalars that collectively break a local U(1) symmetry. These complex scalars interact with fermions through Yukawa interactions that, in addition, satisfy a discrete interchange symmetry, essentially leading to an effective theory below the symmetry breaking scale $M$ involving the inflaton field and two Dirac fermions with a Lagrangian density of the form:
\begin{equation} \label{WLI_lagrangian}
-\mathcal{L}=gM\cos(\phi/M)\bar\psi_1\psi_1+ gM\sin(\phi/M)\bar\psi_2\psi_2~,
\end{equation}
where $g$ is a dimensionless coupling. The particular form of this Lagrangian makes the fermion masses bounded from above, such that large inflaton field values do not lead to heavy fermions, and in addition leads to the cancellation of the leading thermal contributions of the fermion fields to the inflaton's mass. These fermions are also allowed to decay into other light fermions and scalars, not directly coupled to the inflaton, through standard Yukawa interactions parametrized by a dimensionless coupling $h$.

This simple scenario has been shown to lead to a consistent realization of warm inflation with a small number of fields and parameters. The inflaton is, moreover, a gauge singlet, such that the scenario can accommodate different forms of the scalar potential compatible with the reflection symmetry $\phi/M \rightarrow \pi/2-\phi/M$ inherited from the discrete interchange symmetry mentioned above. Moreover, this scenario leads to observational predictions for chaotic inflation with a quartic potential compatible with the latest Planck data \cite{Bastero-Gil:2016qru,Bastero-Gil:2017wwl}, a simple model that is observationally ruled out within the standard ``cold inflation" paradigm.

In this work, we extend the analysis done in \cite{Bastero-Gil:2016qru,Bastero-Gil:2017wwl} for different forms of the scalar potential, thoroughly exploring the parametric regimes where the WLI scenario can be consistently implemented and comparing the associated observational predictions with Planck 2015 data \cite{Planck}. In addition to the quartic monomial potential, we also study a Higgs-like symmetry breaking potential and a non-renormalizable plateau-like potential also considered in \cite{Bastero-Gil:2016mrl, Benetti:2016jhf} in the context of warm inflation. We wish to determine, in particular, the allowed ranges for the dimensionless couplings $g$ and $h$ and the symmetry breaking scale $M$ for which the WLI scenario can be consistently realized with different potentials, as well as characterize the dynamics of warm inflation in the consistent parametric regimes.

This work is organized as follows. In Sect.~\ref{Warm little inflaton 2} we describe the basic dynamics and observational predictions of  warm inflation within the WLI scenario. In Sect.~\ref{Analysis of potentials} we analyze in detail three different potentials: a chaotic quartic potential, a Higgs-like potential and a non-renormalizable plateau-like potential. In all cases, we identify the regions in parameter space where all consistency conditions are satisfied and compute the associated observational predictions. The main conclusions of this work are summarized in Sect.\ref{Conclusions}.


\section{Warm inflation dynamics and primordial perturbation spectrum}\label{Warm little inflaton 2}

In warm inflation the background evolution equations for the inflaton-radiation system are given by: 
\begin{eqnarray}
\ddot{\phi}+(3H+\Upsilon)\dot{\phi}+V'(\phi) &=& 0 \,, \nonumber\\
\dot{\rho}_{R}+4H\rho_{R} &=& \Upsilon\dot{\phi}^{2} \,, \label{background_eqs}
\end{eqnarray}
where dots correspond to time derivatives, primes denote derivatives with respect to $\phi$, $\Upsilon$ is the dissipative coefficient in the leading adiabatic approximation and $H$ is the Hubble parameter, given by the Friedmann equation for a flat FRW universe:
\begin{equation}\label{Hubble}
3H^{2}=\frac{\rho}{M_P^{2}} \,,
\end{equation}
where $\rho=\rho_\phi+\rho_R$ is the total energy density, with $\rho_{\phi}=\dot{\phi}^{2}/2+V(\phi)$. 

In the WLI scenario, the dissipation coefficient resulting from the interaction between the inflaton and the fermions $\psi_1$ and $\psi_2$ in Eq.~(\ref{WLI_lagrangian}) is proportional to the temperature of the radiation bath and given by \cite{Bastero-Gil:2016qru}:
\begin{equation}\label{Upsilon}
\Upsilon=C_{T}T \,, \quad C_{T} \simeq \frac{3g^{2}}{h^{2}(1-0.34\log(h))} \,,
\end{equation}
where $C_{T}$ is a function of the coupling $g$ and the Yukawa coupling $h$ determining the decay of the $\psi_{1,2}$ fermions into a light $\sigma$ scalar and a light $\psi_\sigma$ fermion. The above dissipation coefficient is valid in the high-temperature regime where the fermions are relativistic. Given that, including thermal mass corrections, $m_1^2=g^{2}M^{2}\cos^{2}(\phi/M)+h^{2}T^{2}/8$ and $m_2^2=g^{2}M^{2}\sin^{2}(\phi/M)+h^{2}T^{2}/8$, the fermions remain light during inflation for an arbitrary inflaton value, provided that $gM\lesssim T \lesssim M$. Note that the upper bound on the temperature ensures that the underlying U(1) symmetry is spontaneously broken during inflation.

The contribution of the fermions $\psi_1$ and $\psi_2$ to the finite temperature effective potential is given by \cite{Kapusta:2006,Cline:1997}:
\begin{eqnarray}
V_{T} &\simeq& \left\{-\frac{28\pi^{2}}{15}+h^{2}\left[1+\frac{3h^{2}}{32\pi^{2}}\left[\ln\left(\frac{\mu^{2}}{T^{2}}\right)-c_{f}\right]\right]\right\}\frac{T^{4}}{48}+\left\{1+\frac{3h^{2}}{16\pi^{2}}\left[\ln\left(\frac{\mu^{2}}{T^{2}}\right)-c_{f}\right]\right\}\frac{g^{2}M^{2}}{12}T^{2} \nonumber\\
&& \frac{g^{4}M^{4}}{16\pi^{2}}\left[\cos^{4}(\phi/M)+\sin^{4}(\phi/M) \right]\left[\ln\left(\frac{\mu^{2}}{T^{2}}\right)-c_{f}\right]\,,
\end{eqnarray}
where $\mu$ {\footnote{For convenience throughout this paper we select this scale as the symmetry breaking scale $M$.} is the $\overline{\text{MS}}$ renormalization scale and $c_{f}=2.635$. Its derivatives are given by: 
\begin{eqnarray}
V_{T,\phi} \simeq -\frac{g^{4}M^{3}}{16\pi^{2}}\sin(4\phi/M)\left[\ln\left(\frac{\mu^{2}}{T^{2}}\right)-c_{f}\right]~, \qquad
V_{T,\phi\phi} \simeq -\frac{g^{4}M^{2}}{4\pi^{2}}\cos(4\phi/M)\left[\ln\left(\frac{\mu^{2}}{T^{2}}\right)-c_{f}\right]~, 
\end{eqnarray}
where the leading thermal inflaton mass corrections from both fermions cancel each other, and the remaining oscillatory corrections vanish, on average, for $\phi\gg M$, although we will include them explicitly in our analysis.

The fermion decay width is given by, neglecting the mass of its decay products:
\begin{equation}\label{decay-rate}
\Gamma_{\psi_i}= \frac{h^{2}}{16\pi}\frac{T^{2}m_{i}^{2}}{\omega_{p}^{2}|\mathbf{p}|}\left[F\left(\frac{k_{+}}{T},\frac{\omega_{p}}{T}\right)-F\left(\frac{k_{-}}{T},\frac{\omega_{p}}{T}\right) \right]  \,,
\end{equation}
where $\omega_{p}=\sqrt{m_{i}^{2}+|\mathbf{p}|^{2}}$, $k_{\pm}=(\omega_{p}\pm |\mathbf{p}|)/2$ and 
\begin{equation}
F(x,y)=xy-\frac{x^{2}}{2}+(y-x)\ln\left(\frac{1-e^{-x}}{1+e^{x-y}}\right)+\text{Li}_{2}\left(e^{-x}\right)+\text{Li}_{2}\left(-e^{x-y}\right) \,,
\end{equation}
where $\text{Li}_{2}(z)$ is the dilogarithm function. The thermal mass corrections, given by the Yukawa interactions, dominate over the inflaton contribution to the fermion masses for $h\gg g$ and $T\lesssim M$, such that $ m_{i}^{2}\simeq h^{2}T^{2}/8$. To ensure the validity of the adiabatic approximation in the computation of the dissipation coefficient and that the fermions are in a nearly-thermal equilibrium state, we must then impose $\Gamma_{\psi}/H>1$. For practical purposes, we evaluate the decay width at the momentum value $p_{max}\simeq 3.24T$ that yields the largest contribution to the dissipation coefficient \cite{Bastero-Gil:2016qru}. In addition to this requirement, we demand $T>H$, where a flat space approximation can be employed in the computation of the dissipation coefficient \cite{Landsman}.

The general expression for the amplitude of the primordial curvature power spectrum is given by \cite{Berera:1999ws, Hall:2003zp, Moss:2007cv, Graham:2009bf, Ramos:2013nsa, Bastero-Gil:2014jsa} :
\begin{equation}\label{Spectrum}
\Delta_{\mathcal{R}}^{2}=\frac{V_{*}(1+Q_{*})^{2}}{24\pi^{2}M_P^{4}\epsilon_{*}} \left(1+2n_{*}+\frac{2\sqrt{3}\pi Q_{*}}{\sqrt{3+4\pi Q_{*}}}\frac{T_*}{H_*}\right)G(Q_{*})\,,
\end{equation} 
where all quantities are evaluated when the relevant CMB modes become superhorizon 50-60 e-folds before inflation ends. In the expression above, $n_{*}$ denotes the phase space distribution of inflaton fluctuations at horizon-crossing. Depending on the strength of the interactions between inflaton particles and other particles in the thermal bath (including e.g.~scattering processes), this should interpolate between the Bunch-Davies vacuum, $n_*=0$, and the Bose-Einstein distribution at temperature $T$, $n_*\simeq \left(e^{H_{*}/T_{*}}-1  \right)^{-1}$. We will focus on the latter limiting case in this paper, which we denote as ``thermal" inflaton fluctuations.

The function $G(Q_{*})$ accounts for the growth of inflaton fluctuations due to the coupling to radiation fluctuations through the temperature dependence of the dissipation coefficient and must be determined numerically. This function also exhibits a mild dependence on the form of the scalar potential. Extending the analysis in \cite{Bastero-Gil:2016qru} for the potentials considered in this work, we find:
\begin{eqnarray}
G(Q_{*}) &\simeq & 1+0.0185 \, Q_{*}^{2.315}+0.335 \, Q_{*}^{1.364}  \,,\quad \text{quartic potential,} \\
               &           & 1+0.01 \, Q_{*}^{1.8}+0.18 \, Q_{*}^{1.4}  \,,\quad \text{Higgs-like and plateau-like potentials} 
\end{eqnarray}
For thermalised inflation fluctuations, $n_{*}\simeq T_{*}/H_{*}\gtrsim 1$, and since $T_{*}/H_{*}=3Q_{*}/C_{T}$ the resulting dimensionless power spectrum has the form: 
\begin{equation}\label{SpectrumQ}
\Delta_{\mathcal{R}}^{2}=\frac{5C_{T}^{4}}{72\pi^{4}g_{*}}Q_{*}^{-3}\left(1+\frac{6}{C_{T}}Q_{*}+\frac{\sqrt{3}\pi}{\sqrt{3+4\pi Q_{*}}}\frac{6}{C_{T}}Q_{*}^{2} \right)G(Q_{*})\,.
\end{equation} 
This implies that the measured amplitude of the primordial power spectrum $\Delta_{\mathcal{R}}^{2}\simeq 2.2\times10^{-9}$ constrains the observationally consistent scenarios independently of the form of the scalar potential. In particular, it yields an upper bound $C_{T}\lesssim0.02$, which implies that the coupling $g$ must be at least an order of magnitude smaller than the coupling $h$.

From the amplitude of the curvature power spectrum, we may determine the scalar spectral index  $n_{s}-1\simeq d\ln\Delta_{\mathcal{R}}^{2}/dN_{e}$, which we may write as:
\begin{equation}\label{ns}
n_{s}=1+\frac{Q_{*}}{3+5Q_{*}}\left(6\epsilon_{\phi}-2\eta_{\phi}\right)\frac{d\ln\Delta_{\mathcal{R}}^{2}}{dQ_{*}} \,.
\end{equation}
Since, for $T\ll M_P$, gravitational waves are not significantly affected by thermal effects, the primordial tensor spectrum is given by the standard inflationary form $\Delta_{t}^{2}=2H_{*}^{2}/(\pi^{2}M_P^{2})$. The tensor-to-scalar ratio $r=\Delta_{\mathcal{R}}^{2}/\Delta_{t}^{2}$ is nevertheless affected, and as mentioned above typically reduced, by the modifications to the scalar curvature perturbations introduced by dissipation.


\section{Analysis of different inflaton potentials}\label{Analysis of potentials}

Previous studies of the WLI scenario have shown that it may be consistently implemented and yield observationally viable predictions with a quartic inflaton potential for specific parameter values \cite{Bastero-Gil:2016qru,Bastero-Gil:2017wwl}. In this section, we will perform a full parametric analysis of this case, considering all dynamical consistency conditions, and then extend the study to additional typical forms of the inflaton potential. We will require, in particular, that the fermions remain light during inflation and that they maintain a near-equilibrium distribution, such that dissipative processes also occur in an adiabatic regime. This implies imposing the conditions $\Gamma_{\psi}/H>1$, $T/H>1$ and $gM\lesssim T \lesssim M$ for 50-60 e-folds of inflationary expansion. After determining the regions of parameter space where these conditions are satisfied, we will then compute the corresponding inflationary observables. 

Since the coupling constants involved in the dynamics differ by several orders of magnitude, for instance  $\lambda\sim10^{-14}$ for the inflaton self-interactions and $h=\mathcal{O}(1)$, it is useful to use rescaled quantities in the numerical procedure to evolve the background field and radiation equations (\ref{background_eqs}). In particular, in the numerical code we use $\tilde{C}_{T}=\tilde{g}^{2}\lambda^{1/4}/f(\tilde{h})$, $f(\tilde{h})=3/(\tilde{h}^{2}(1-0.34\log(\tilde{h})))$, and $\tilde{g}=g/\lambda^{1/4}$. This allows us to consider input values $g_0$ and $h_0$ for the couplings and a reference input value $\lambda_0=10^{-14}$ for the inflaton self-coupling. With these input values we evolve the background equations from initial conditions yielding a large number of e-folds ($>60$), and then determine the values of the different dynamical quantities ($\phi$, $Q$) at horizon-crossing 50-60 e-folds before the end of inflation. From this we then determine the physical value of $\lambda$ yielding the measured amplitude of the scalar curvatue power spectrum, and use the rescaled quantities above to compute the physical values of the couplings $g$ and $h$.


\subsection{Chaotic inflation with a quartic potential: $V(\phi)=\frac{\lambda}{4}\phi^{4}$}\label{Quartic}

For the quartic potential, we show in Fig.~\ref{fig:quartic_h-M-g} the regions in the $(g,M/M_P)$ plane where all dynamical consistency conditions are satisfied for two different values of the Yukawa coupling, $h=2$ and $h=3$ ($h\simeq h_0$ and $g\simeq g_0$ in this case), indicating the regions where each condition fails.

\begin{figure}[htbp] 
\includegraphics[scale=0.41]{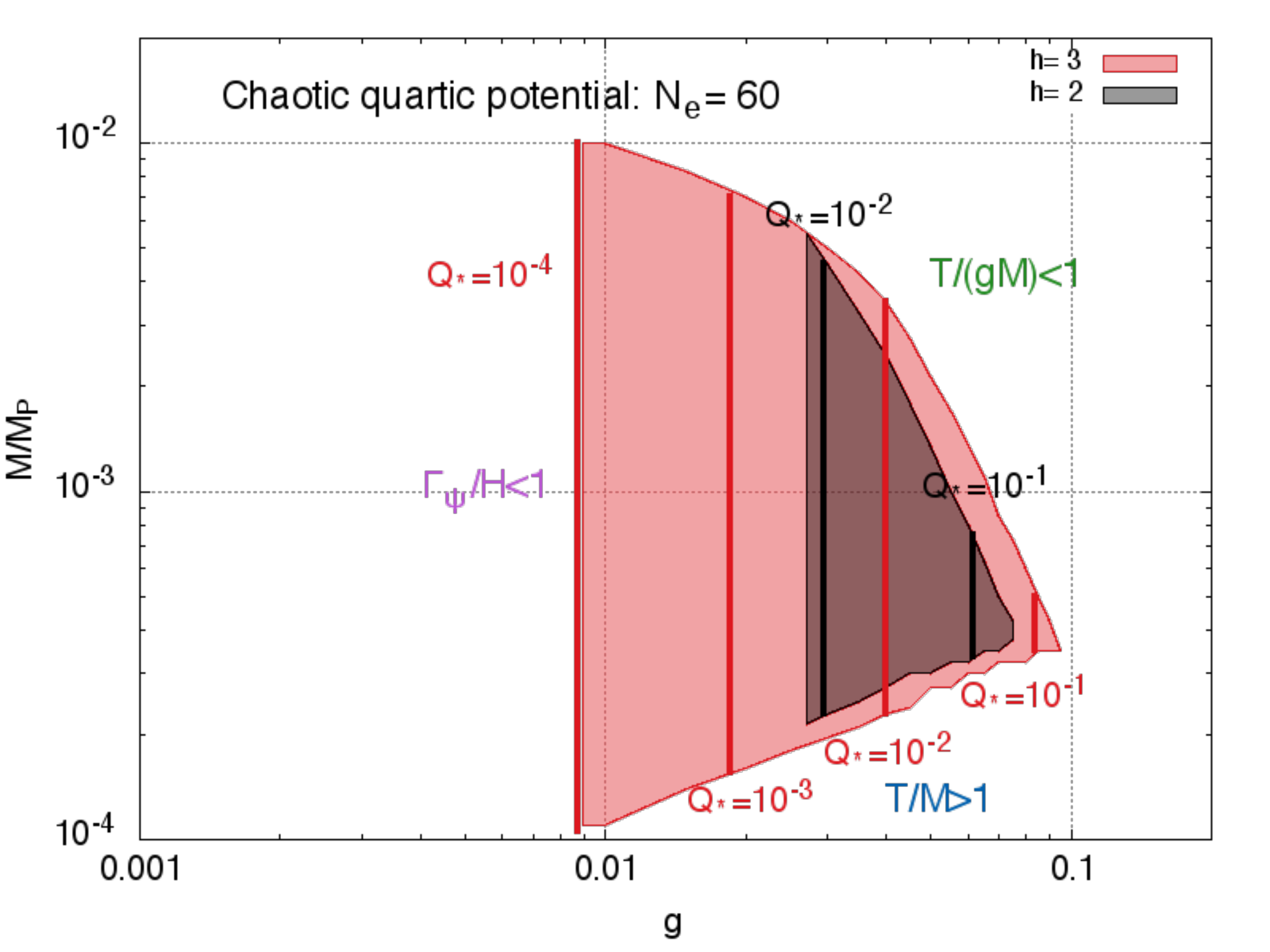}\vspace{0.5cm}
\includegraphics[scale=0.41]{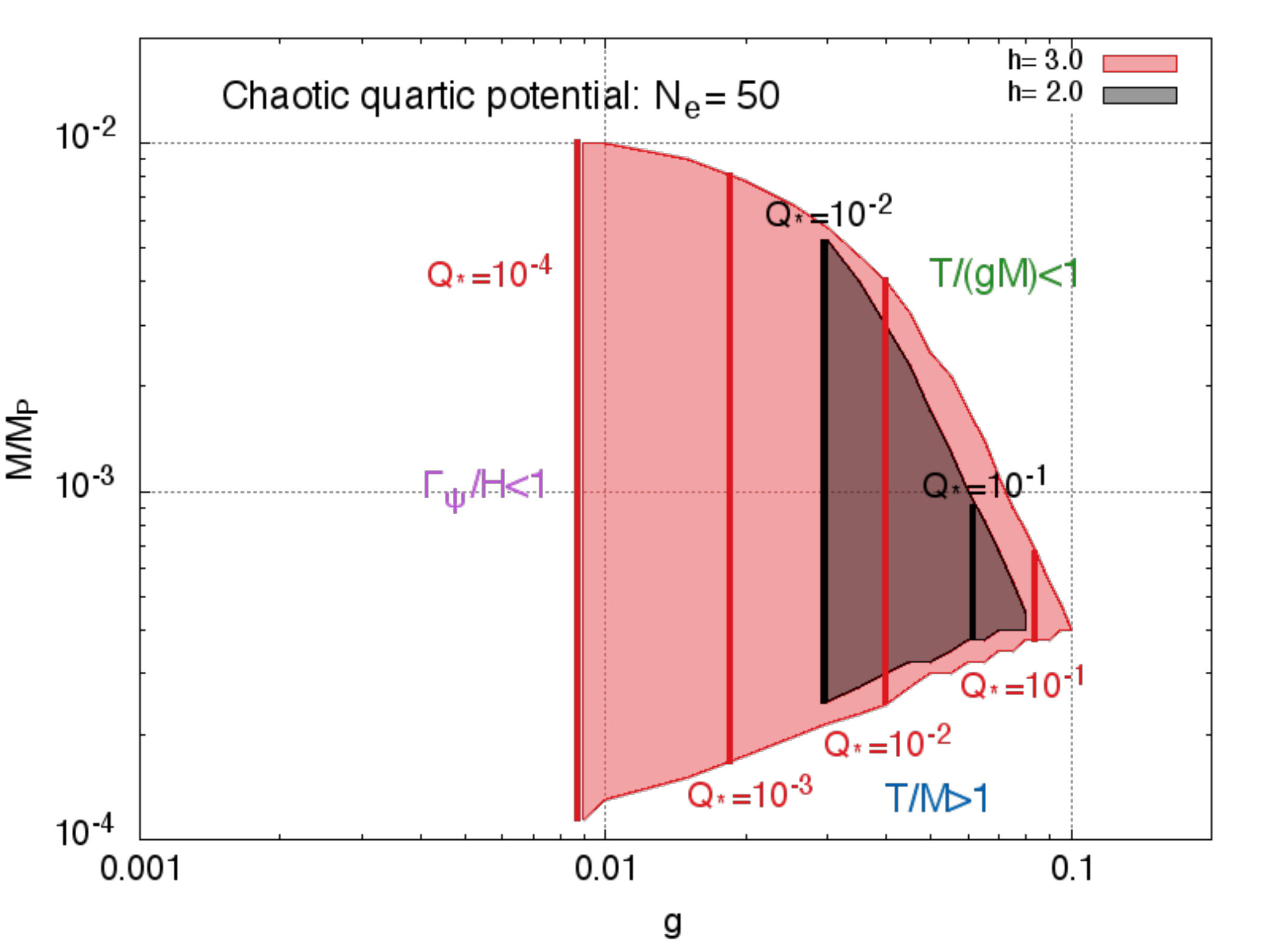}
\caption{Allowed regions in the plane $(g, M/M_P)$ for the chaotic quartic potential with $h=2$ (gray) and $h=3$ (red), for $N_e=50$ (right) and $N_e=60$ (left). Notice that there are no allowed regions for $h\simeq 1$. The vertical lines correspond to different values of the dissipative ratio at horizon-crossing, $Q_*$.}\label{fig:quartic_h-M-g}
\end{figure}

 As one can easily see, the adiabatic condition $\Gamma_\psi >H$ implies a lower bound on $g\gtrsim 0.01$, while the conditions on the temperature limit this coupling from above, $g\lesssim 0.1$. We find no consistent solutions for $h\lesssim 1$ (due to the condition $\Gamma_\psi>H$), with larger values of this coupling increasing the range of the allowed values for $g$ and $M$. Note, however, that for larger values of the Yukawa coupling perturbation theory may break down. The symmetry breaking scale may take values in the range $10^{-4} M_P - 10^{-2} M_P$.  Detailed limits are given in Table \ref{Table:Quartic} in the appendix.

We also find that the dissipative ratio at horizon-crossing can consistently take values in the range $10^{-4}\lesssim Q_*\lesssim10^{-1}$, the lower bound being set by the condition of nearly-thermalized fermions and the upper bound by the high-temperature approximation. Thus, generically we find that inflation must start in the weak dissipative regime, although $Q$ increases during inflation for the quartic potential so that in a wide region of parameter space one reaches $Q>1$ before the end of inflation, a necessary condition for radiation to dominate after the slow-roll regime with no further reheating (see Eq.~(\ref{radiation_abundance})).

In Fig.~\ref{fig:quartic_obs_all} we show the predictions for the scalar spectral index and tensor-to-scalar ratio in the allowed parametric ranges, exhibiting a remarkable consistency with the Planck data. This is particularly relevant given that the quartic potential is already excluded by Planck data within the cold inflation paradigm. The tensor-to-scalar ratio lies in the range $10^{-3}\lesssim r \lesssim 10^{-2}$, with a smaller Yukawa coupling suppressing the amount of tensor modes. 


\begin{figure}[h] 
\centering\includegraphics[scale=0.5]{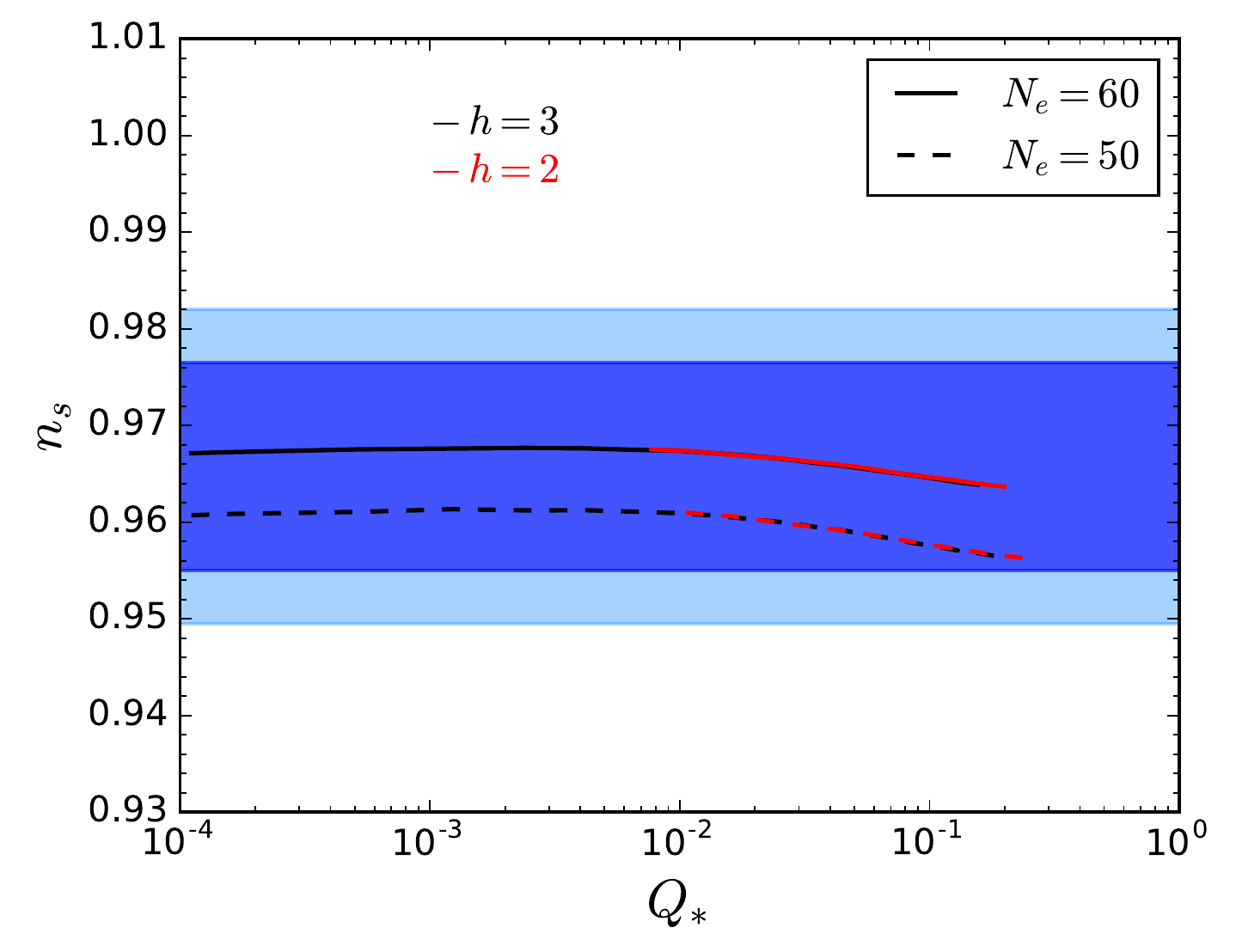}\vspace{0.5cm}
\centering\includegraphics[scale=0.5]{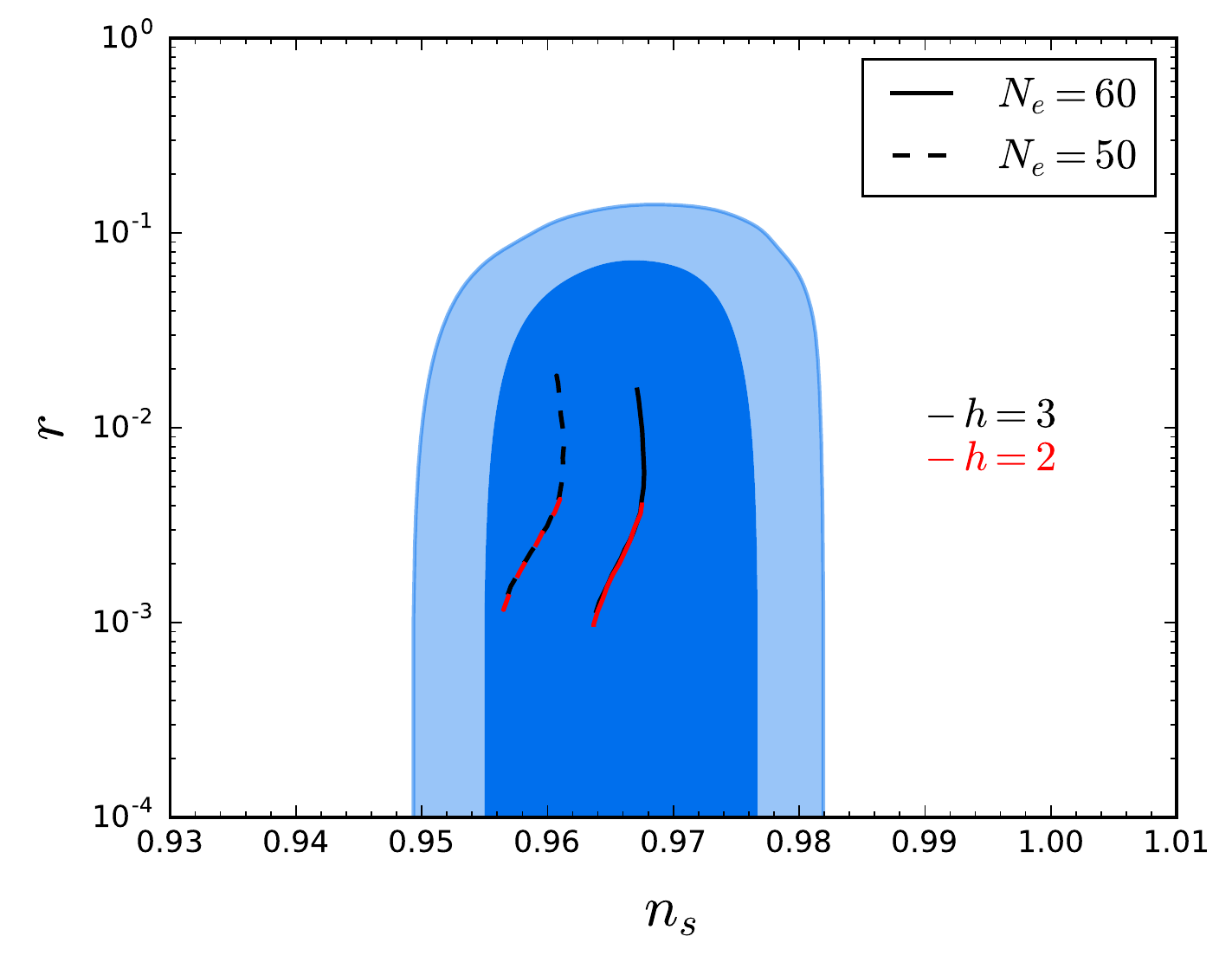}
\caption{Observational predictions of the WLI scenario with a quartic potential for 50-60 e-folds of inflation and two distinct values of the Yukawa coupling $h=2$ and $h=3$. The plot on the left shows the spectral index $n_{s}$ as a function of the dissipative ratio at horizon-crossing, $Q_*$, while the plot on the right shows the allowed trajectories in the $(n_s,r)$ plane. The blue contours correspond to the $68\%$ and $95\%$ C.L. results from Planck 2015 TT+lowP data \cite{Planck}.}\label{fig:quartic_obs_all} 
\end{figure}

The agreement between the WLI quartic model and the Planck data had already been observed in \cite{Bastero-Gil:2016qru,Bastero-Gil:2017wwl}, although without taking into account all the dynamical consistency conditions. Interestingly, these conditions exclude the parametric regimes for which $Q_*\gtrsim 1$, where the growing mode due to the coupling between inflaton and radiation perturbations would render the spectrum more blue-tilted and disfavored by Planck. Hence, it is truly remarkable that consistency of the analysis leads to a full agreement with Planck and also to a finite range for the tensor-to-scalar ratio within the reach of B-mode polarization experiments in the near future (see e.g.~\cite{Kamionkowski:2015yta}).


\subsection{Inflation with a Higgs-like potential: $V(\phi)=\frac{\lambda}{4}(\phi^{2}-v^{2})^{2}$}\label{Higgs}

For a Higgs-like ``mexican hat" inflaton potential, with inflation occurring in the ``hilltop" part of the potential, we show in Fig.~\ref{fig:higgs_Ne_60-M-g} the regions in the $(g, M/M_P)$ plane where all dynamical consistency conditions are satisfied, for different values of the symmetry breaking scale $v$ and two different values of the input Yukawa coupling $h_0$ (which differs from the physical value of the coupling as discussed below). We fix $N_e=60$ for clarity in this case, with the results being very similar for $N_e=50$.
 
\begin{figure}[h!] 
\centering\includegraphics[scale=0.41]{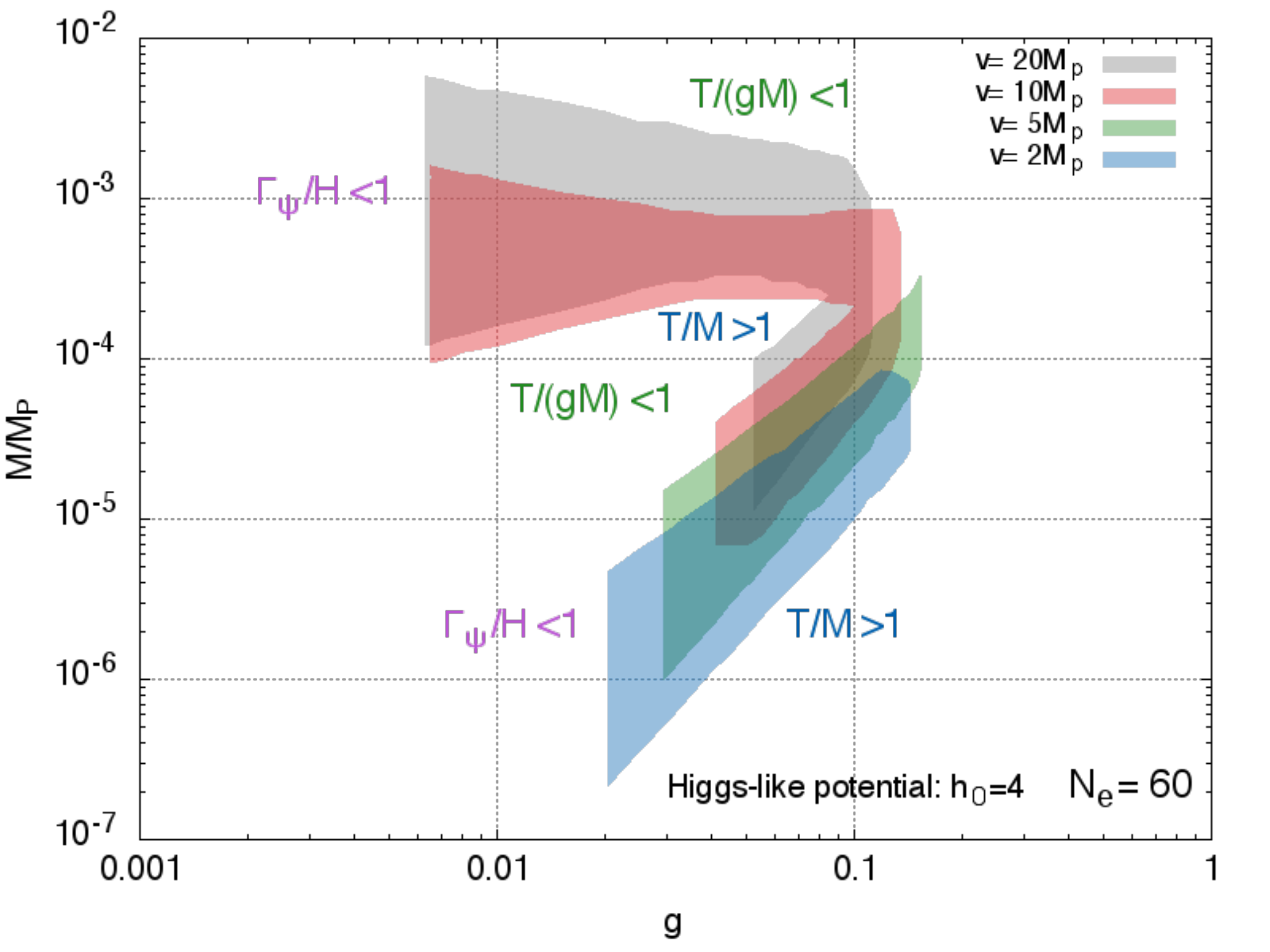}\vspace{0.5cm}
\centering\includegraphics[scale=0.41]{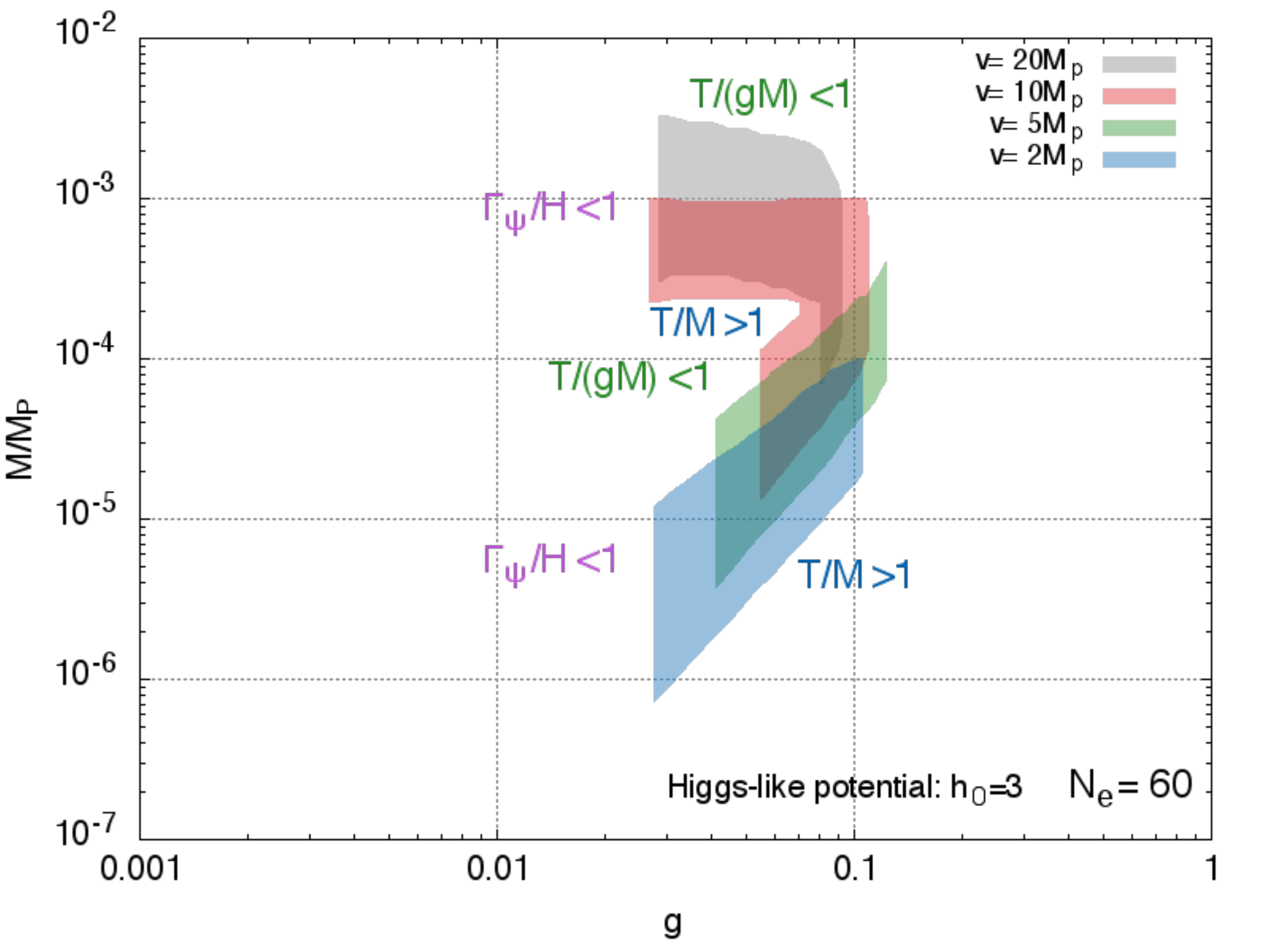}\vspace{-0.5cm}
\caption{Allowed regions in the $(g, M/M_P)$ plane for the Higgs-like potential with $h_0=3$ (right) and $h_0=4$ (left), for $N_e=60$ and different values of the symmetry breaking scale $v$. Notice that there are no allowed regions for $h_0\simeq 2$.}\label{fig:higgs_Ne_60-M-g}
\end{figure}

As one can see in Fig.~\ref{fig:higgs_Ne_60-M-g}, larger values of $v$ imply larger values of the WLI symmetry breaking scale $M$, which may lie in a wider range than for the quartic potential studied earlier. As for the latter, the condition $\Gamma_\psi>H$ yields a lower bound on the coupling $g\gtrsim 0.01$, while the conditions for light fermions and U(1) symmetry breaking yield an upper bound $g\lesssim 0.1$. We also find that the input value for the Yukawa coupling $h_0\gtrsim 2$ in order to satisfy all consistency conditions, with larger values of $h_0$ increasing the allowed region in the $(g, M/M_P)$ plane. 

Detailed limits on the parameters are given in Table \ref{Table:Higgs_Ne_60} in the appendix, where one can see that the WLI scale $10^{-7}M_P\lesssim M\lesssim 10^{-2} M_P$ and that, despite the lower bound on $h_0$, the physical value of the Yukawa coupling can take values $h\lesssim 1$. In fact, we find that in this regime inflation occurs in the strong dissipation regime already at horizon-crossing, being consistent to have $Q_* \lesssim 10^3$. To better understand the different dynamical regimes at horizon-crossing, we plot in Fig.~(\ref{fig:higgs_Ne_60_g-h-Q}) the relation between the physical couplings $g$ and $h$ and $Q_*$ for $N_e=60$.

\begin{figure}[htbp] 
\centering\includegraphics[scale=0.9]{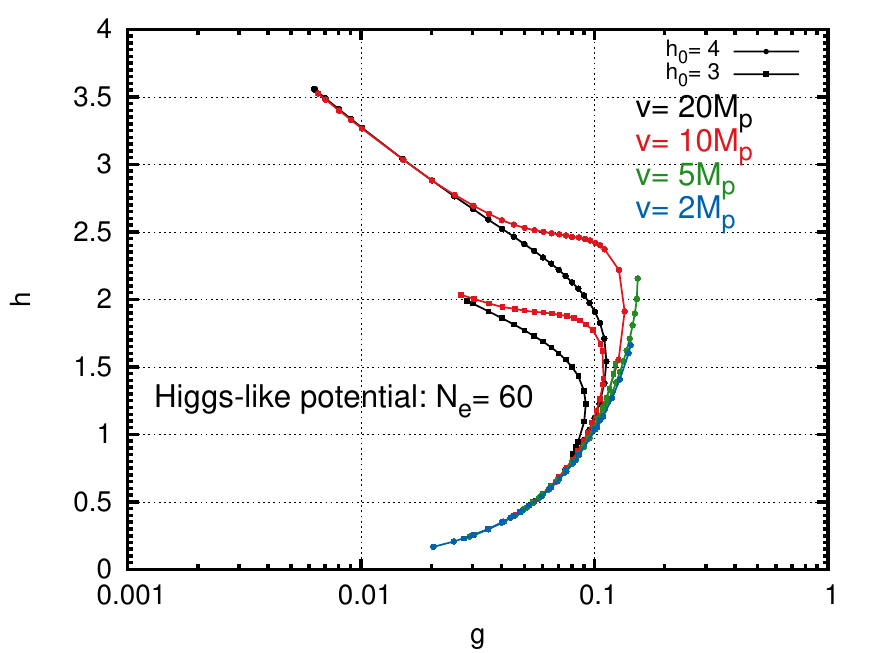}\vspace{0.5cm}
\centering\includegraphics[scale=0.9]{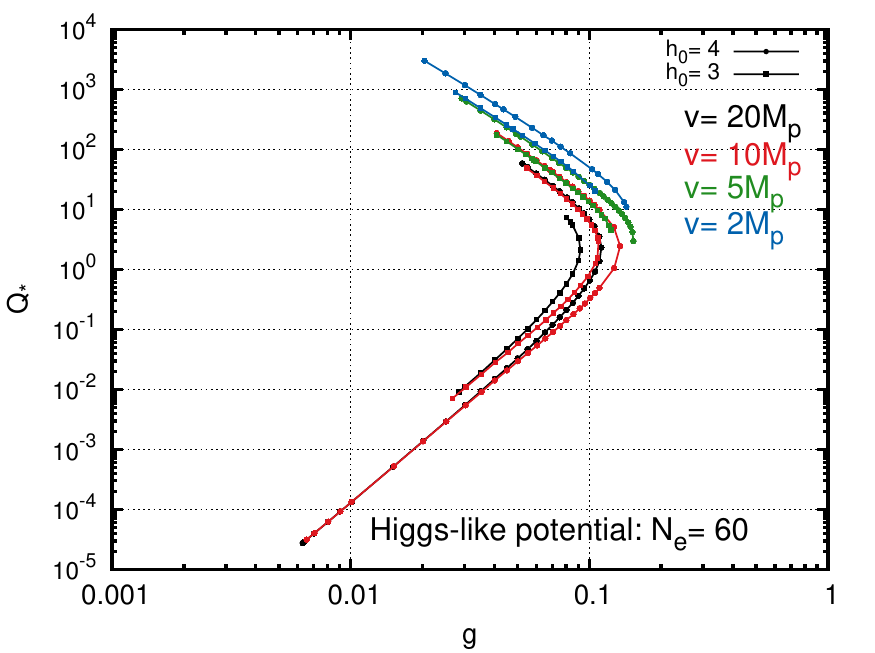}\vspace{-0.5cm}
\caption{Allowed values of the physical couplings $g$ and $h$ (left) in the WLI scenario with a Higgs-like potential for $N_e=60$ and two different values of the input parameter $h_0$, for distinct values of the inflaton potential minimum $v$. The corresponding values of $Q_*$ are also plotted as a function of $g$ (right).}\label{fig:higgs_Ne_60_g-h-Q}
\end{figure}      

In Fig.~\ref{fig:higgs_Ne_60_g-h-Q} we can clearly identify two distinct allowed parametric regimes. First, a regime where $h\gtrsim 1$ decreases with $g$ that yields weak dissipation at horizon-crossing, much like for the quartic potential. Second, a regime where $h\lesssim 1$ increases with $g$ where dissipation is already strong at horizon-crossing, $Q_*\gtrsim 1$, with $Q_*$ increasing with decreasing values of $h$. The latter constitutes an allowed parametric window that was absent for the quartic potential.

Despite the larger parametric window found for the Higgs-like potential, 
in comparison with the quartic potential, we need to investigate the 
observational consistency of the allowed parametric ranges, particularly 
in the regime $Q_*\gtrsim 1$. The obtained inflationary observables are 
shown in Fig.~\ref{fig:higgs_ns-Qstar-r}.
 
\begin{figure}[htbp] 
\centering\includegraphics[scale=0.5]{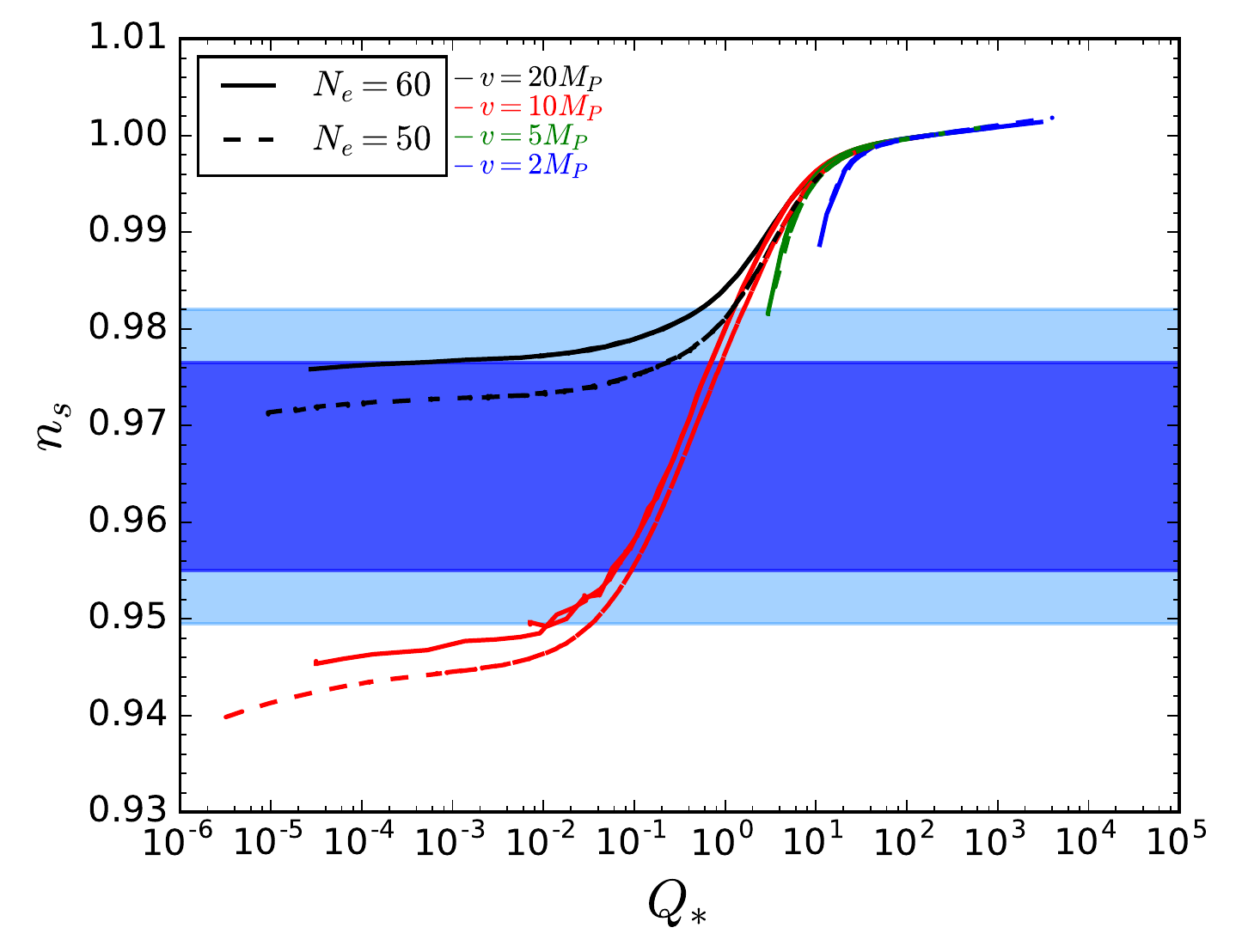}\vspace{0.5cm}
\centering\includegraphics[scale=0.5]{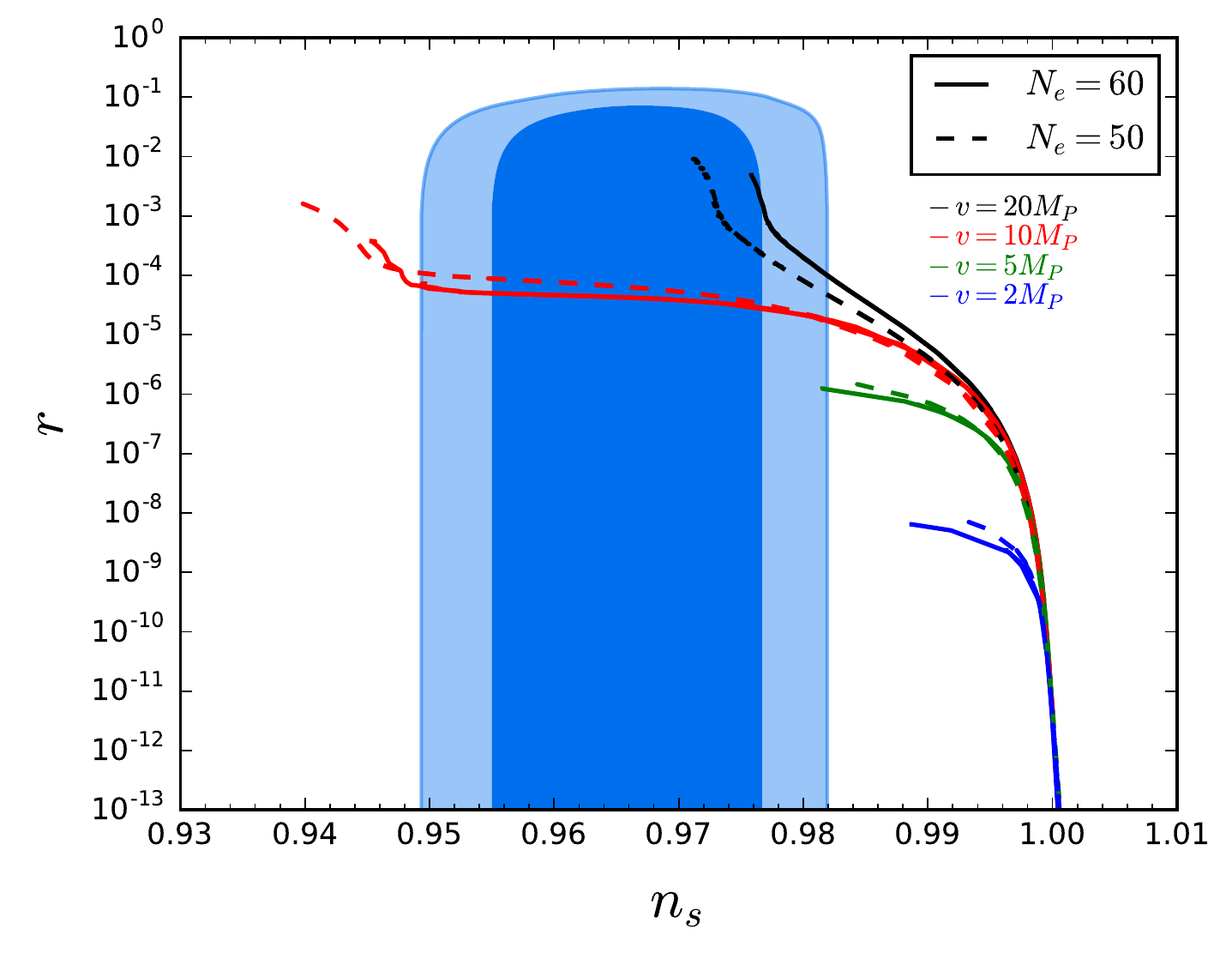}
\caption{Observational predictions of the WLI scenario with a Higgs-like potential for 50-60 e-folds of inflation and different values of the symmetry breaking scale $v$. The plot on the left shows the spectral index $n_{s}$ as a function of the dissipative ratio at horizon-crossing, $Q_*$, while the plot on the right shows the allowed trajectories in the $(n_s,r)$ plane. The blue contours correspond to the $68\%$ and $95\%$ C.L. results from Planck 2015 TT+lowP data \cite{Planck}.}\label{fig:higgs_ns-Qstar-r}
\end{figure}

It is clear in Fig.~\ref{fig:higgs_ns-Qstar-r} that the parametric regime where $Q_*\gtrsim 1$ (and $h\lesssim 1$) is disfavoured by the Planck data, essentially due to the growing mode in the spectrum of inflaton perturbations associated with their coupling to radiation perturbations and which makes the spectrum more blue-tilted with increasing $Q_*$. Hence, in tune with what we found earlier for the quartic model, this scenario is only observationally viable for $Q_*\lesssim 1$ and values of the Yukawa coupling $h\gtrsim 1$, although we emphasize that $Q$ is dynamical and a strong dissipative regime can be attained before the end of inflation. Values for which $Q_* \gtrsim 1$ correspond to the region $M \lesssim 10^{-4}M_P$, and therefore, like for the quartic model, agreement with Planck restricts the symmetry breaking scale to the range $10^{-4} M_P \lesssim M \lesssim 10^{-2} M_P$. In addition, observations also restrict the Higgs symmetry breaking scale to the range $5M_P\lesssim v\lesssim 20M_P$, which is essentially a restriction on the $\eta_\phi$ slow-roll parameter at horizon-crossing. An important difference between the Higgs-like potential and the quartic potential is that the tensor-to-scalar ratio can be much lower in the latter, with $r\gtrsim 10^{-6}$ for the Higgs-like potential in the allowed window.


\subsection{Inflation with a nonrenormalizable plateau-like potential: $V(\phi)=\lambda v^{4} \left(1-3\frac{\phi^{4}}{v^{4}}+2\frac{\phi^{6}}{v^{6}}\right)\,.$}\label{Hilltop}

To complete our discussion of inflationary potentials, we consider a non-renormalizable plateau-like potential with quartic and sextic inflaton monomials. This has a symmetry-breaking shape like the Higgs-like potential studied above, with the important difference that both slow-roll parameters $\epsilon_\phi$ and $\eta_\phi$ vanish at the origin, making the resulting inflationary plateau much flatter. This potential is, in fact, very similar to the Coleman-Weinberg potential typically considered in several inflationary models (see e.g.~\cite{Bastero-Gil:2016mrl}).

In Fig.~\ref{fig:hilltop_Ne_60-M-g} we show the regions in the $(g, M/M_P)$ plane where all dynamical consistency conditions are satisfied, for different values of the symmetry breaking scale $v$ and two different values of the input Yukawa coupling $h_0$ (which differs from the physical value of the coupling). As for the Higgs-like potential, we fix $N_e=60$ for clarity of the plots, with the results being very similar for $N_e=50$.
   
\begin{figure}[h] 
\centering\includegraphics[scale=0.41]{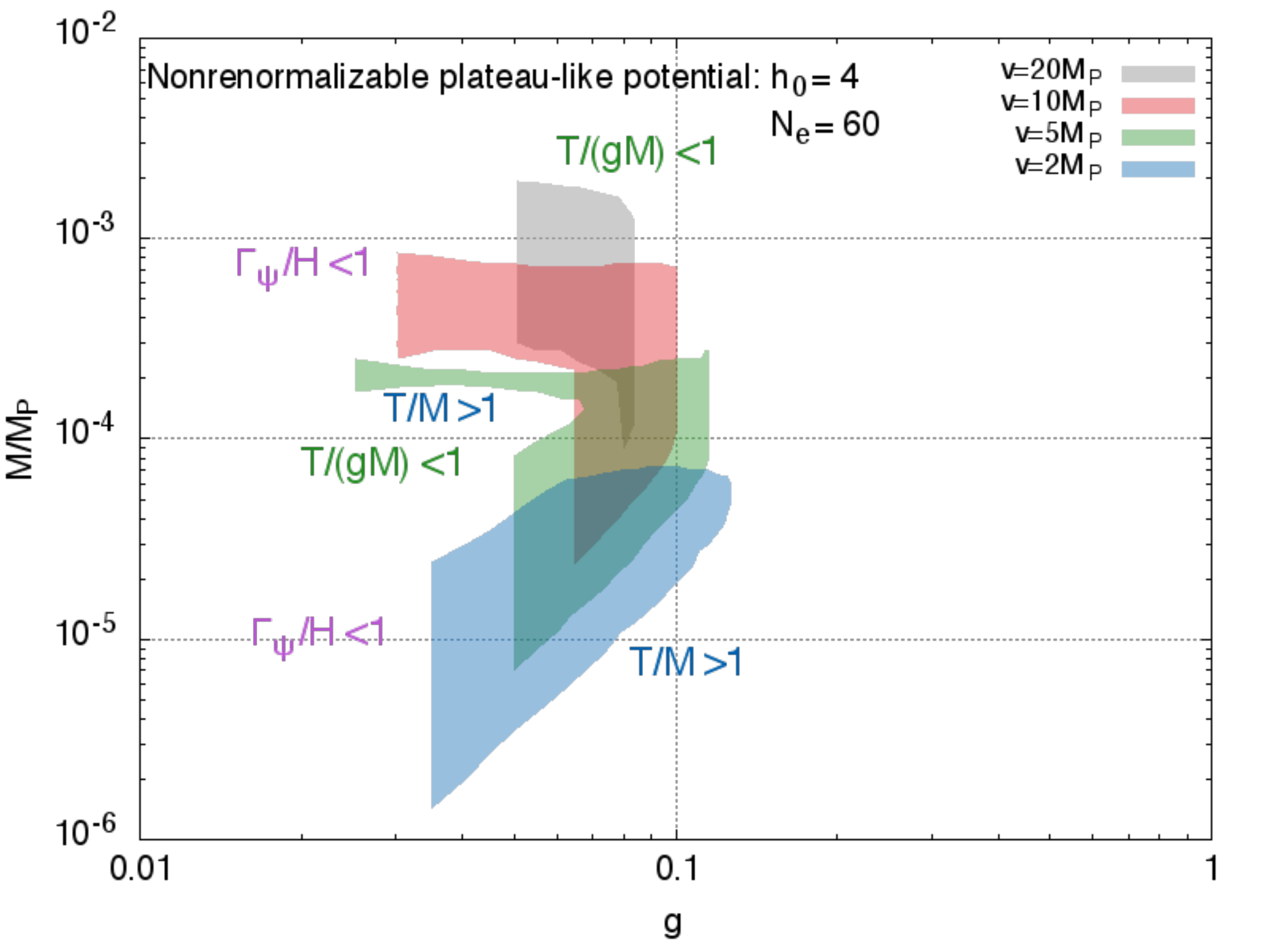}\vspace{0.5cm}
\centering\includegraphics[scale=0.41]{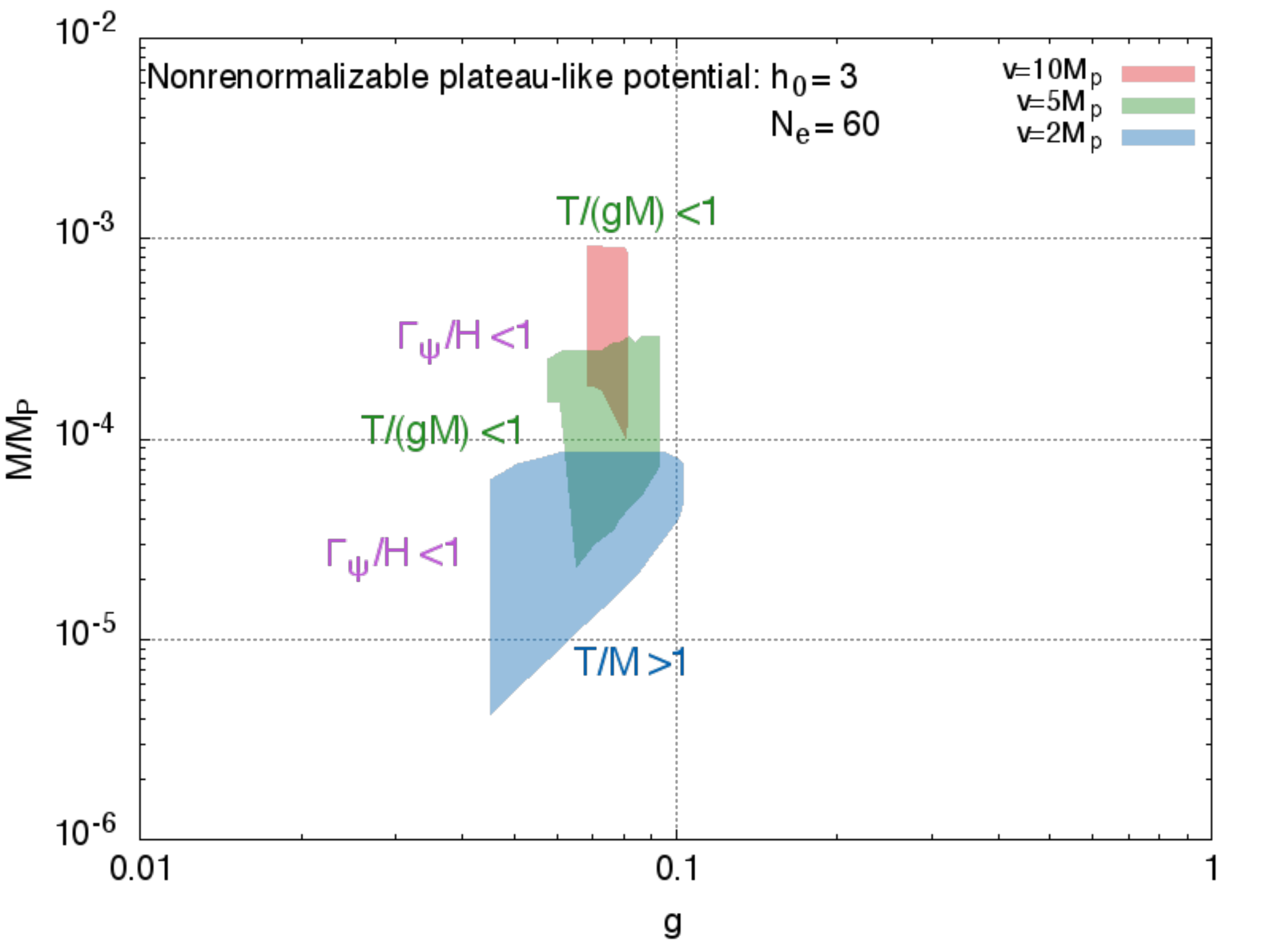}
\caption{Allowed regions in the plane $(g, M/M_P)$ for the nonrenormalizable plateau-like potential with $h_0=3$ (right) and $h_0=4$ (left), for $N_e=60$ and different values of the symmetry breaking scale $v$. Notice that there are no allowed regions for $h_0\simeq 2$.}\label{fig:hilltop_Ne_60-M-g}
\end{figure}

Fig.~\ref{fig:hilltop_Ne_60-M-g} shows that the allowed parametric regions for the nonrenormalizable plateau-like potential are very similar to the Higgs-like potential, with larger values of $v$ shifting the allowed window for the symmetry breaking scale $M$ towards larger values, and larger values of $h_0$ enhancing the allowed parametric window. This window is somewhat narrower than for the Higgs-like potential, but again we find a lower bound $h_0\gtrsim 2$ on the input value of the Yukawa coupling, and that $0.01\lesssim g \lesssim 0.1$. In this case the decrease in the allowed parametric region between $h_0=4$ and $h_0=3$ is more pronounced than for the Higgs-like potential, and in fact the case $v=20M_P$ is excluded by the dynamical consistency conditions for $h_0=3$. Detailed limits on the parameters are given in Table~\ref{Table:Hilltop_Ne_60} in the appendix.
  
For the plateau-like potential, we thus find that $10^{-6}\lesssim M/M_P\lesssim 10^{-3}$, and as for the Higgs-like potential we also obtain an allowed region where $h\lesssim 1$ and $Q_*\gtrsim 1$. The dissipative ratio may also attain values $Q_*\lesssim 10^3$ in this case, although somewhat smaller than for the Higgs-like scenario. The relation between the physical couplings and $Q_*$ is qualitatively analogous to the Higgs-like potential as shown in Fig.~\ref{fig:hilltop_Ne_60_g-h-Q}.

\begin{figure}[htbp] 
\centering\includegraphics[scale=0.9]{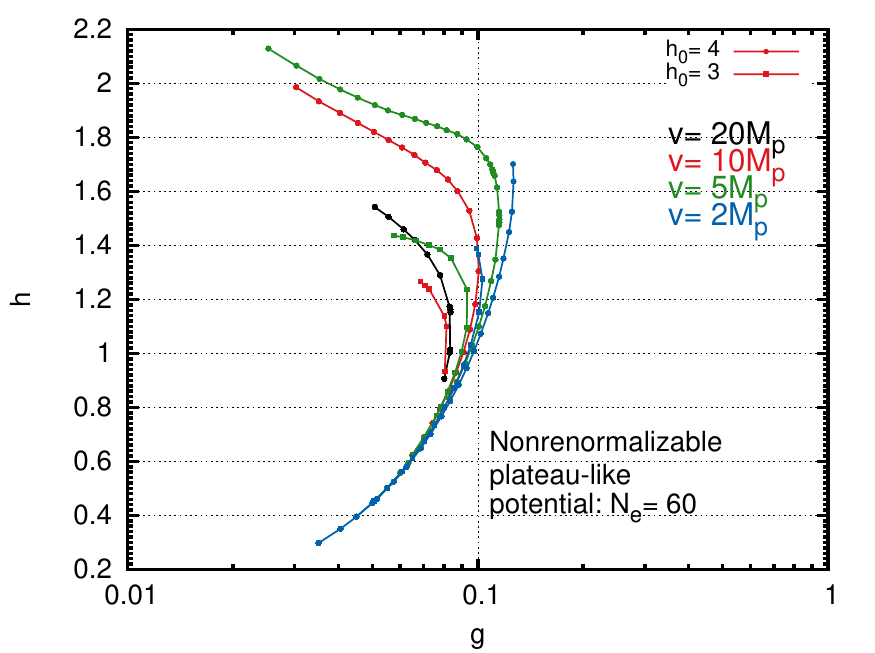}\vspace{0.5cm}
\centering\includegraphics[scale=0.9]{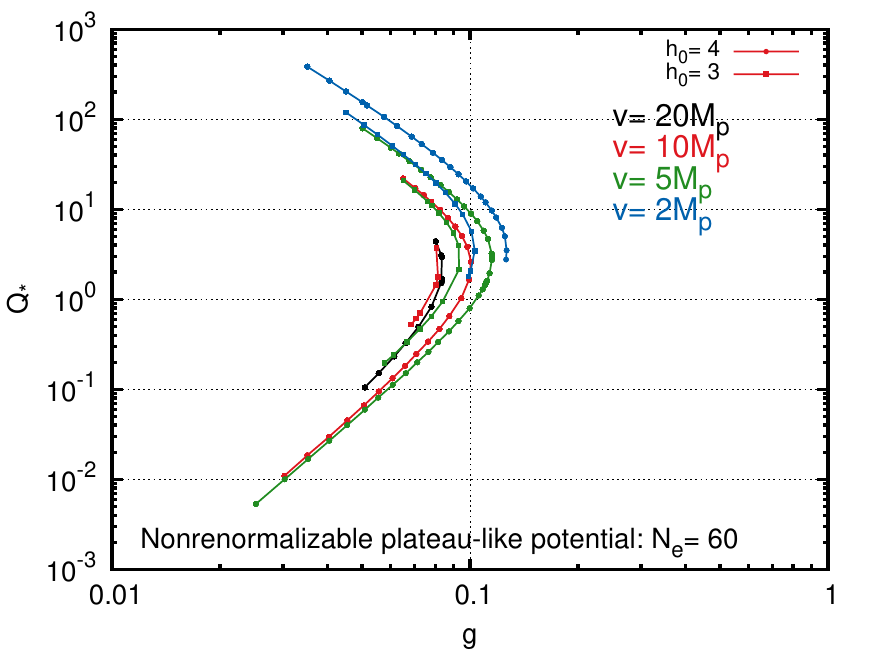}
\caption{Allowed values of the physical couplings $g$ and $h$ (left) in the WLI scenario with a nonrenormalizable plateau-like potential for $N_e=60$ and two different values of the input parameter $h_0$, for distinct values of the inflaton minimum $v$. The corresponding values of $Q_*$ are also plotted as a function of $g$ (right).}\label{fig:hilltop_Ne_60_g-h-Q}
\end{figure}

Not surprisingly, scenarios with strong dissipation at horizon-crossing are observationally disfavoured by Planck data, again due to the growing mode in the power spectrum, as shown in Fig.~\ref{fig:hilltop_ns-Qstar-r}. Nevertheless, we may have viable scenarios with $Q_*\sim 3$ in this case, which is somewhat larger than for the other potentials studied in this work, and also for $v=2M_P$, which did not occur for the Higgs-like potential. Of the three forms of the potential considered in this work, this flatter plateau is thus the one that allows for stronger dissipation at horizon-crossing and hence larger values of the inflaton mass compared to the Hubble parameter, showing that warm inflation can significantly alleviate the ``eta-problem". The non-renormalizable plateau also yields lower allowed values for the tensor-to-scalar ratio, with $r\gtrsim 10^{-8}$ in this case.

\begin{figure}[htbp] 
\centering\includegraphics[scale=0.5]{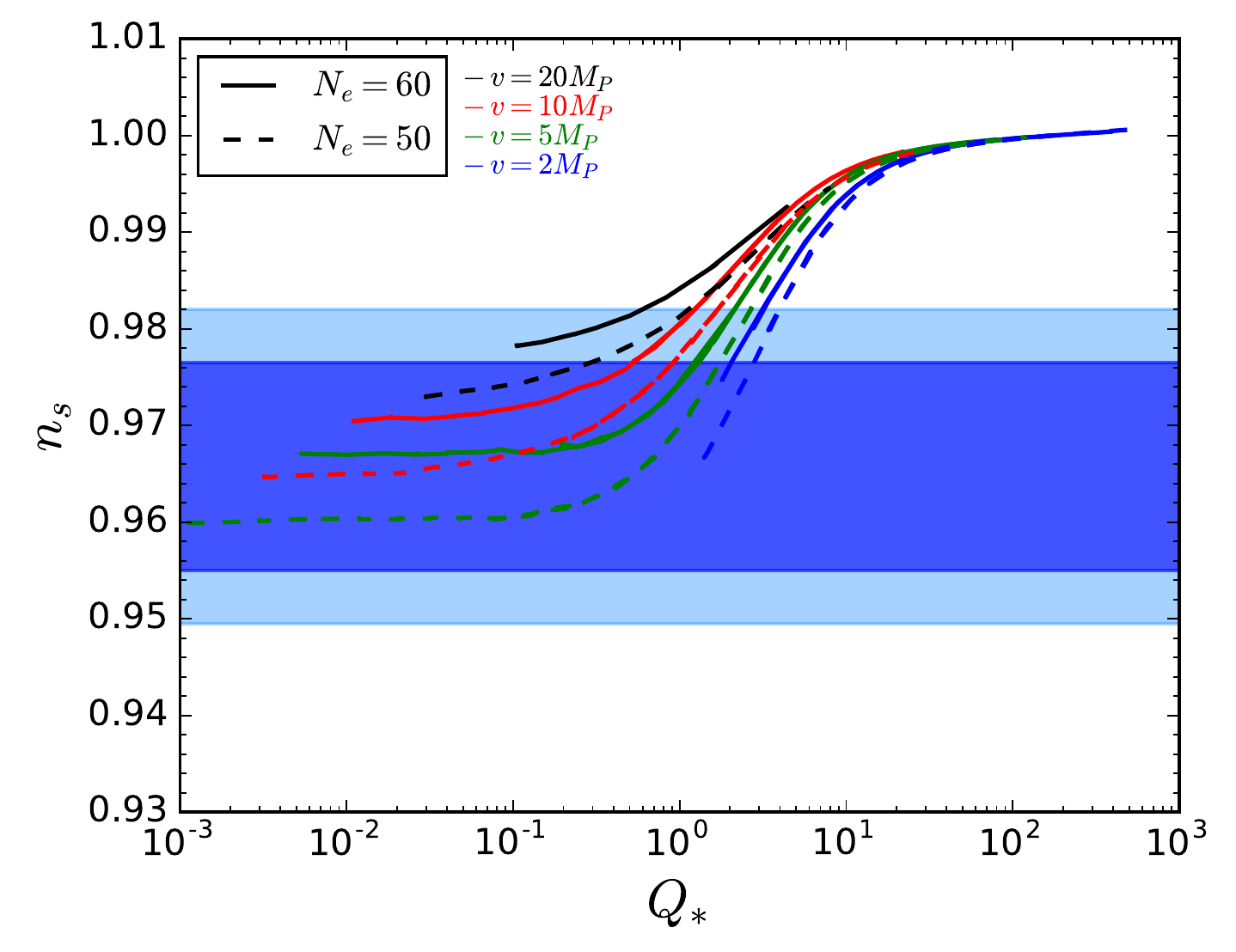}\vspace{0.5cm}
\centering\includegraphics[scale=0.5]{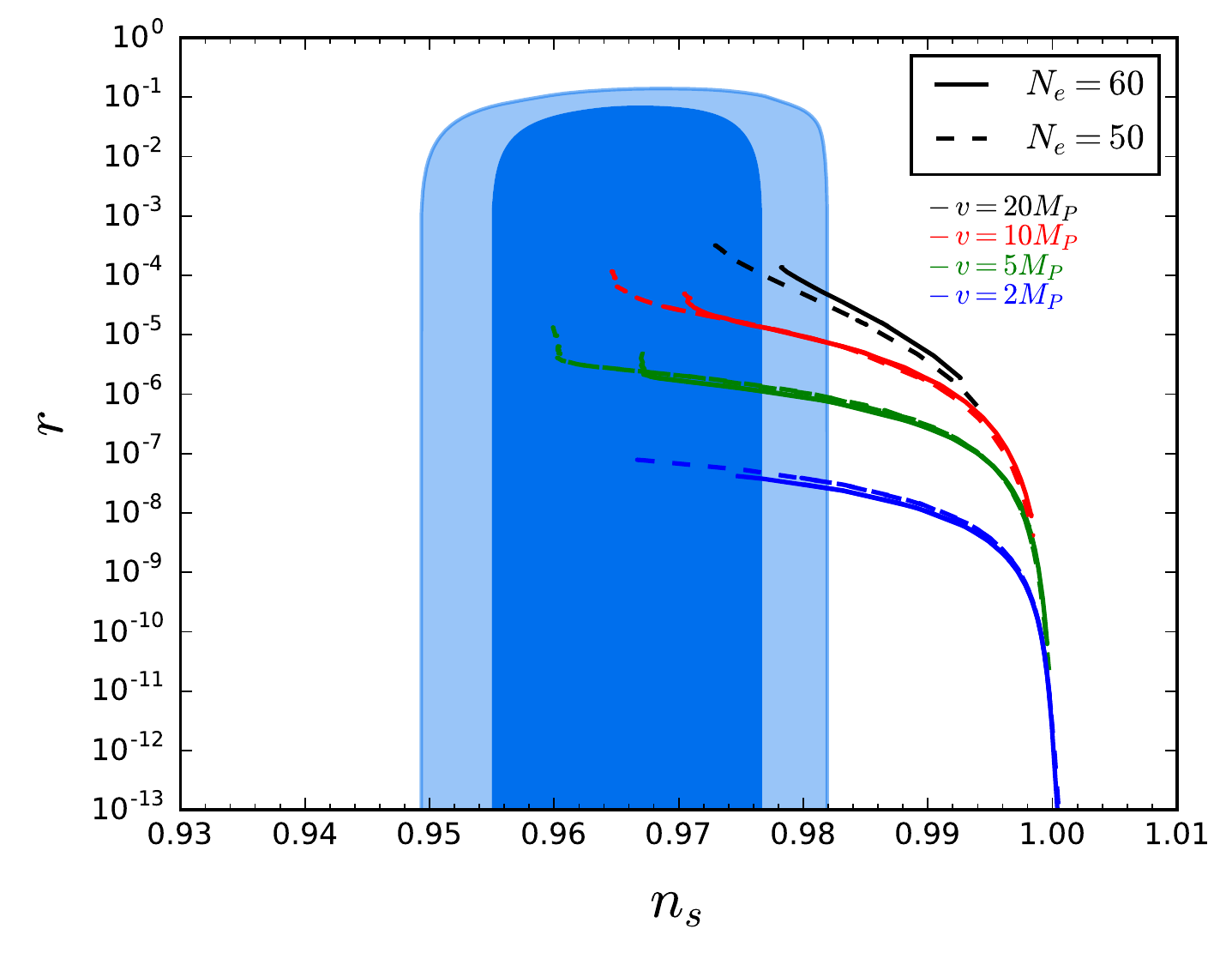}
\caption{Observational predictions of the WLI scenario with a nonrenormalizable plateau-like potential for 50-60 e-folds of inflation and different values of the symmetry breaking scale $v$. The plot on the left shows the spectral index $n_{s}$ as a function of the dissipative ratio at horizon-crossing, $Q_*$, while the plot on the right shows the allowed trajectories in the $(n_s,r)$ plane. The blue contours correspond to the $68\%$ and $95\%$ C.L. results from Planck 2015 TT+lowP data \cite{Planck}.}\label{fig:hilltop_ns-Qstar-r}
\end{figure}


\section{Conclusions}\label{Conclusions}

The {\it Warm Little Inflaton} scenario \cite{Bastero-Gil:2016qru} is currently the most promising realization of warm inflation within a quantum field theory model, involving only a handful of fields and employing symmetries to cancel the troublesome thermal corrections to the inflaton potential that have frustrated earlier attempts to construct a simple warm inflation model in the high-temperature regime. Despite this crucial cancellation, a consistent inflationary dynamics requires that several conditions are satisfied throughout inflation, in particular that the fermions coupled directly to the inflaton remain light and in a near-thermal equilibrium state, ensuring also that the dissipation coefficient can be computed in the adiabatic approximation and neglecting spacetime curvature corrections. 

In this work, we have studied the implementation of this scenario for different forms of the inflaton potential, considering a large-field chaotic model and small-field models like the Higgs-like and non-renormalizable plateau-like potentials. We have scanned the parameter space of the model imposing all the relevant consistency conditions and compared the associated observational predictions with the most recent Planck data. We have included in our analysis all thermal corrections to the effective scalar potential that remain upon cancellation of the leading inflaton thermal mass corrections.

Although results are quantitatively different for the three potentials, we have found several generic features of the WLI model. Consistent realizations require a coupling between the inflaton and fermion fields $0.01\lesssim g \lesssim 0.1$ and that the latter decay through a Yukawa coupling $h\gtrsim 1$ (although within the perturbative regime). This ensures that the fermions can remain light throughout inflation ($gM \lesssim T$) for temperatures below the symmetry breaking scale $T\lesssim M$, and that they decay faster than the inflationary expansion. Imposing these conditions we have found that the condition for a warm inflationary regime, $T\gtrsim H$, is easily satisfied, not imposing any additional constraints. Dynamical consistency also imposes $10^{12}\ \mathrm{GeV}\lesssim M \lesssim 10^{16}\ \mathrm{GeV}$, with observations favoring the larger values, thus suggesting a possible connection between the WLI scenario and grand unification theories.    

For the quartic chaotic scenario, dynamical consistency imposes $Q_*\lesssim 0.1$ for the dissipative ratio at horizon-crossing, while for the two hilltop potentials dynamical constraints allow for $Q_*\lesssim 10^3$. However, we have found that all scenarios with $Q_*\gtrsim 1$ are disfavored by the Planck data, essentially due to a growing mode in the curvature power spectrum associated with the coupling between inflaton and radiation fluctuations that is present in all warm inflation scenarios with $\Upsilon\propto T^n$, $n>0$, and which generically leads to larger $n_s$ values. Nevertheless, in all cases discussed we may reach values of $Q_*$ close or even slightly above unity, and since $Q$ grows throughout inflation for all potentials considered, we have thus obtained scenarios where the dynamics of warm inflation occurs essentially in the strong dissipation regime, $Q>1$, and for which the ``eta-problem" can thus be significantly alleviated.

The dynamical and observational constraints (essentially on the scalar spectral index $n_s$) also allow us to place lower limits on the tensor-to-scalar ratio ranging from $r\gtrsim 10^{-3}$ for the quartic model down to $r\gtrsim 10^{-8}$ for the plateau-like potential, and in all cases consistent models have $r\lesssim 10^{-2}$. In the chaotic model the dynamical constraints are, in fact, sufficient to limit the values of $n_s$ and $r$ to values within the Planck bounds, with $10^{-3}\lesssim r \lesssim 10^{-2}$ being potentially within the reach of ongoing and planned  B-mode polarization experiments \cite{Kamionkowski:2015yta}.

The remarkable agreement of the quartic potential with Planck data is particularly significant given that such self-interactions should generically be present in the scalar potential, not being forbidden by any symmetries\footnote{The interchange symmetry actually requires the potential to be of the form $V(\phi)=\lambda M^4(\phi/M-\pi/4)^4$, but this is well approximated by a quartic monomial for $\phi\gg M$.}, and dominate over a quadratic mass term for sufficiently large field values. This scenario thus yields the most natural renormalizable realization of chaotic inflation, where slow-roll is a phase-space attractor as opposed to plateau-like models (see e.g.~\cite{Brandenberger:2016uzh}). While in the absence of dissipation its predictions are in severe tension with observational data \cite{Planck}, it is in perfect agreement with Planck within the WLI realization of warm inflation (see also \cite{Bartrum:2013fia}). 

Of course that, as in cold inflation models, several scalar potentials will be excluded by observational data within the WLI scenario. We have focused, in this work, on three types of potential for which we find agreement with observations in a broad region of parameter space, thus showing that the WLI scenario can lead to different realizations of warm inflation that are both dynamically and observationally consistent. The modifications to the primordial perturbation spectrum induced by dissipative and thermal effects within the WLI scenario generically lead to a more blue-tilted scalar spectrum and a more suppressed tensor component with respect to cold inflation models with the same potential functions. Hence, scalar potentials that were excluded within the cold inflation paradigm for being too red-tilted and yielding a too large tensor-to-scalar ratio are generically in better agreement with observations within warm inflation.

We thus hope that this work motivates further studies of the WLI scenario and of other simple realizations of warm inflation, both from the model-building and from the observational perspectives, considering also its impact on the subsequent cosmic history, towards building a complete and testable particle physics description of the early Universe.

\vspace{0.5cm}

\acknowledgments
RHJ acknowledges CONACyT for financial support.  AB is supported by STFC. MBG is partially supported by ``Junta de Andalucia" (FQM101) and the University of Granada (PP2015-03). JGR is supported by the FCT Investigator Grant No.~IF/01597/2015 and partially by the H2020-MSCA-RISE-2015 Grant No. StronGrHEP-690904 and by the CIDMA Project No.~UID/MAT/04106/2013.


\appendix

\section{Detailed parameter limits for different potentials}

\begin{table}[htbp]
\begin{center}
\begin{tabular}{|c|cccccc|}
\hline
 $N_{e}=60$ & $M_{min}/M_P$ & $M_{max}/M_P$ & $g_{min}$ & $g_{max}$ & $Q_{* min}$ & $Q_{* max}$ \\
\hline
h=3 & $1.1\times10^{-4}$ & $1\times10^{-2}$ & $8.9\times10^{-3}$ & $0.095$ & $1\times10^{-4}$ & $0.16$  \\
h=2 & $2.15\times10^{-4}$ & $5.5\times10^{-3}$ & $0.027$ & $0.075$ & $7.7\times10^{-3}$ & $0.2$  \\
\hline
\hline
 $N_{e}=50$ & $M_{min}/M_P$ & $M_{max}/M_P$ & $g_{min}$ & $g_{max}$ & $Q_{* min}$ & $Q_{* max}$ \\
 \hline
 h=3 & $1.15\times10^{-4}$ & $1\times10^{-2}$ & $8.9\times10^{-3}$ & $0.1$ & $1.1\times10^{-4}$ & $0.19$  \\
h=2 & $2.5\times10^{-4}$ & $5.25\times10^{-3}$ & $0.03$ & $0.08$ & $0.01$ & $0.25$  \\
\hline
\hline								
\end{tabular}
\end{center}
\caption{\label{Table:Quartic} Allowed parametric ranges for the WLI scenario with a quartic potential for $N_e=50,60$ and $h=2,3$.}
\end{table}


\newpage
\begin{table}[htbp]
\begin{center}
$N_e=60$\\\vspace{0.1cm}
\begin{tabular}{|c|cccccccc|}
\hline
 $h_{0}=4$ & $M_{min}/M_P$ & $M_{max}/M_P$ & $h_{min}$ & $h_{max}$ &  $g_{min}$ & $g_{max}$ & $Q_{* min}$ & $Q_{* max}$ \\
\hline
$v/M_P=20$ & $1.15\times10^{-5}$ & $5.75\times10^{-3}$ & $0.475$ & $3.559$ & $6.3\times10^{-3}$ & $0.112$ & $2.76\times10^{-5}$ & $58.237$  \\
$v/M_P=10$ & $7.0\times10^{-6}$ & $1.6\times10^{-3}$ & $0.354$ & $3.526$ & $6.54\times10^{-3}$ & $0.135$ & $3.135\times10^{-5}$ & $186.515$  \\
$v/M_P=5$   & $1.0\times10^{-6}$ & $3.25\times10^{-4}$ & $0.243$ & $2.154$ & $0.029$ & $0.153$ & $2.95$ & $704.09$  \\
$v/M_P=2$   & $2.15\times10^{-7}$ & $8.5\times10^{-5}$ & $0.167$ & $1.66$ & $0.02$ & $0.143$ & $11.0$ & $2999.9$  \\
\hline
\hline
 $h_{0}=3$ & $M_{min}/M_P$ & $M_{max}/M_P$ & $h_{min}$ & $h_{max}$ &  $g_{min}$ & $g_{max}$ & $Q_{* min}$ & $Q_{* max}$ \\
\hline
$v/M_P=20$ & $7.0\times10^{-5}$ & $3.25\times10^{-3}$ & $0.856$ & $1.99$ & $0.028$ & $0.092$ & $9.11\times10^{-3}$ & $7.32$  \\
$v/M_P=10$ & $1.35\times10^{-5}$ & $1.0\times10^{-3}$ & $0.499$ & $2.035$ & $0.027$ & $0.106$ & $7.14\times10^{-3}$ & $48.4$  \\
$v/M_P=5$   & $3.75\times10^{-6}$ & $4.0\times10^{-4}$ & $0.35$ & $1.52$ & $0.041$ & $0.123$ & $4.5$ & $174.75$  \\
$v/M_P=2$   & $7.25\times10^{-7}$ & $1.0\times10^{-4}$ & $0.229$ & $1.112$ & $0.028$ & $0.106$ & $20.18$ & $883.8$  \\
\hline
\hline
\end{tabular} \\\vspace{0.3cm}$N_e=50$\\\vspace{0.1cm}
\begin{tabular}{|c|cccccccc|}
\hline
 $h_{0}=4$ & $M_{min}/M_P$ & $M_{max}/M_P$ & $h_{min}$ & $h_{max}$ &  $g_{min}$ & $g_{max}$ & $Q_{* min}$ & $Q_{* max}$ \\
\hline
$v/M_P=20$ & $1.15\times10^{-5}$ & $7.25\times10^{-3}$ & $0.475$ & $4.04$ & $4.75\times10^{-3}$ & $0.117$ & $9.29\times10^{-6}$ & $65.55$  \\
$v/M_P=10$ & $3.75\times10^{-6}$ & $2.75\times10^{-3}$ & $0.354$ & $4.61$ & $3.62\times10^{-3}$ & $0.141$ & $3.19\times10^{-6}$ & $209.381$  \\
$v/M_P=5$   & $9.25\times10^{-7}$ & $3.75\times10^{-4}$ & $0.241$ & $2.233$ & $0.029$ & $0.162$ & $3.55$ & $805.51$  \\
$v/M_P=2$   & $1.75\times10^{-7}$ & $9.75\times10^{-5}$ & $0.154$ & $1.689$ & $0.019$ & $0.147$ & $14.763$ & $3983.59$  \\
\hline
\hline
 $h_{0}=3$ & $M_{min}/M_P$ & $M_{max}/M_P$ & $h_{min}$ & $h_{max}$ &  $g_{min}$ & $g_{max}$ & $Q_{* min}$ & $Q_{* max}$ \\
\hline
$v/M_P=20$ & $4.5\times10^{-5}$ & $4.25\times10^{-3}$ & $0.694$ & $2.272$ & $0.02$ & $0.096$ & $2.46\times10^{-3}$ & $14.714$  \\
$v/M_P=10$ & $1.3\times10^{-5}$ & $1.7\times10^{-3}$ & $0.5$ & $2.577$ & $0.016$ & $0.115$ & $8.6\times10^{-4}$ & $54.289$  \\
$v/M_P=5$   & $3.75\times10^{-6}$ & $4.0\times10^{-4}$ & $0.354$ & $1.555$ & $0.041$ & $0.128$ & $5.256$ & $195.866$  \\
$v/M_P=2$   & $7.25\times10^{-7}$ & $1.2\times10^{-4}$ & $0.228$ & $1.120$ & $0.027$ & $0.107$ & $24.48$ & $991.673$  \\
\hline
\hline
\end{tabular}
\end{center}
\caption{\label{Table:Higgs_Ne_60} Allowed parametric ranges for the WLI scenario with a Higgs-like potential for $N_e=60$ (top) and $N_e=50$ (bottom), with $h_0=3,4$ and for different values of the symmetry breaking scale $v$.}
\end{table}


\begin{table}[h]
\begin{center}
$N_e=60$\\\vspace{0.1cm}
\begin{tabular}{|c|cccccccc|}
\hline
 $h_{0}=4$ & $M_{min}/M_P$ & $M_{max}/M_P$ & $h_{min}$ & $h_{max}$ &  $g_{min}$ & $g_{max}$ & $Q_{* min}$ & $Q_{* max}$ \\
\hline
$v/M_P=20$ & $9.0\times10^{-5}$ & $1.95\times10^{-3}$ & $0.906$ & $1.542$ & $0.051$ & $0.084$ & $0.105$ & $4.413$  \\
$v/M_P=10$ & $2.4\times10^{-5}$ & $8.5\times10^{-4}$ & $0.623$ & $1.986$ & $0.030$ & $0.101$ & $0.011$ & $22.042$  \\
$v/M_P=5$   & $7.0\times10^{-6}$ & $2.75\times10^{-4}$ & $0.449$ & $2.129$ & $0.025$ & $0.115$ & $5.33\times10^{-3}$ & $79.60$  \\
$v/M_P=2$   & $1.45\times10^{-6}$ & $7.25\times10^{-5}$ & $0.298$ & $1.701$ & $0.035$ & $0.127$ & $2.776$ & $385.681$  \\
\hline
\hline
 $h_{0}=3$ & $M_{min}/M_P$ & $M_{max}/M_P$ & $h_{min}$ & $h_{max}$ &  $g_{min}$ & $g_{max}$ & $Q_{* min}$ & $Q_{* max}$ \\
\hline
$v/M_P=20$ & $--$ & $--$ & $--$ & $--$ & $--$ & $--$ & $--$ & $--$  \\
$v/M_P=10$ & $1.0\times10^{-4}$ & $9.25\times10^{-4}$ & $0.932$ & $1.266$ & $0.069$ & $0.081$ & $0.526$ & $3.726$  \\
$v/M_P=5$   & $2.3\times10^{-5}$ & $3.25\times10^{-4}$ & $0.625$ & $1.436$ & $0.058$ & $0.093$ & $0.196$ & $20.894$  \\
$v/M_P=2$   & $4.25\times10^{-6}$ & $8.5\times10^{-5}$ & $0.396$ & $1.389$ & $0.045$ & $0.103$ & $1.799$ & $119.033$  \\
\hline
\hline
\end{tabular} \\\vspace{0.3cm}$N_e=50$\\\vspace{0.1cm}
\begin{tabular}{|c|cccccccc|}
\hline
 $h_{0}=4$ & $M_{min}/M_P$ & $M_{max}/M_P$ & $h_{min}$ & $h_{max}$ &  $g_{min}$ & $g_{max}$ & $Q_{* min}$ & $Q_{* max}$ \\
\hline
$v/M_P=20$ & $6.75\times10^{-5}$ & $2.45\times10^{-3}$ & $0.786$ & $1.764$ & $0.038$ & $0.088$ & $0.029$ & $7.876$  \\
$v/M_P=10$ & $2.1\times10^{-5}$ & $1.15\times10^{-3}$ & $0.584$ & $2.284$ & $0.022$ & $0.105$ & $3.2\times10^{-3}$ & $29.013$  \\
$v/M_P=5$   & $5.5\times10^{-6}$ & $3.5\times10^{-4}$ & $0.396$ & $2.514$ & $0.017$ & $0.121$ & $1.125\times10^{-3}$ & $118.615$  \\
$v/M_P=2$   & $1.1\times10^{-6}$ & $8.5\times10^{-5}$ & $0.256$ & $1.771$ & $0.031$ & $0.134$ & $3.409$ & $603.231$  \\
\hline
\hline
 $h_{0}=3$ & $M_{min}/M_P$ & $M_{max}/M_P$ & $h_{min}$ & $h_{max}$ &  $g_{min}$ & $g_{max}$ & $Q_{* min}$ & $Q_{* max}$ \\
\hline
$v/M_P=20$ & $--$ & $--$ & $--$ & $--$ & $--$ & $--$ & $--$ & $--$  \\
$v/M_P=10$ & $8.0\times10^{-5}$ & $1.0\times10^{-3}$ & $0.879$ & $1.427$ & $0.060$ & $0.086$ & $0.23$ & $5.404$  \\
$v/M_P=5$   & $1.9\times10^{-5}$ & $4.0\times10^{-4}$ & $0.570$ & $1.602$ & $0.048$ & $0.099$ & $0.077$ & $29.578$  \\
$v/M_P=2$   & $3.5\times10^{-6}$ & $1.0\times10^{-4}$ & $0.357$ & $1.532$ & $0.041$ & $0.110$ & $1.404$ & $169.071$  \\
\hline
\hline
\end{tabular}
\end{center}
\caption{\label{Table:Hilltop_Ne_60} Allowed parametric ranges for the WLI scenario with a nonrenormalizable plateau-like potential for $N_e=60$ (top) and $N_e=50$ (bottom), with $h_0=3,4$ and for different values of the symmetry breaking scale $v$.}
\end{table}




\newpage
\begin{thebibliography}{99}

\bibitem{inflation}
 A.~A.~Starobinsky,
  Phys.\ Lett.\ B {\bf 91}, 99 (1980);
  K.~Sato,
  Mon.\ Not.\ Roy.\ Astron.\ Soc.\  {\bf 195}, 467 (1981);
  A.~H.~Guth,
  Phys.\ Rev.\  {\bf D23}, 347 (1981);
  A.~Albrecht, P.~J.~Steinhardt,
  Phys.\ Rev.\ Lett.\  {\bf 48}, 1220 (1982);
  A.~D.~Linde,
  Phys.\ Lett.\  {\bf B108}, 389 (1982). 
  
\bibitem{Graham:2008vu} 
  I.~G.~Moss and C.~M.~Graham,
  Phys.\ Rev.\ D {\bf 78}, 123526 (2008)
  [arXiv:0810.2039 [hep-ph]].

\bibitem{Berera:1995wh} 
  A.~Berera and L.~-Z.~Fang,
  Phys.\ Rev.\ Lett.\  {\bf 74}, 1912 (1995) 
  [astro-ph/9501024].
  
  
\bibitem{Berera:1995ie} 
  A.~Berera,
  Phys.\ Rev.\ Lett.\  {\bf 75}, 3218 (1995)
  [astro-ph/9509049].

\bibitem{Berera:2008ar} 
  A.~Berera, I.~G.~Moss and R.~O.~Ramos,
  Rept.\ Prog.\ Phys.\  {\bf 72}, 026901 (2009)
  [arXiv:0808.1855 [hep-ph]].
  
  \bibitem{Berera:1996fm}
 A.~Berera,
 Phys.\ Rev.\ D {\bf 55}, 3346 (1997)
 [hep-ph/9612239].  
  
\bibitem{Berera:1999ws}
 A.~Berera,
 Nucl.\ Phys.\ B {\bf 585}, 666 (2000)
 [hep-ph/9904409].


\bibitem{BasteroGil:2009ec} 
  M.~Bastero-Gil and A.~Berera,
  Int.\ J.\ Mod.\ Phys.\ A {\bf 24}, 2207 (2009)
  [arXiv:0902.0521 [hep-ph]].
  
\bibitem{Graef:2018ulg} 
  L.~L.~Graef and R.~O.~Ramos,
  arXiv:1805.05985 [gr-qc].
  
\bibitem{Gron:2018rtj} 
  O.~Gron,
  Universe {\bf 4}, no. 2, 15 (2018).

\bibitem{Oyvind Gron:2016zhz} 
  O.~Gron,
  Universe {\bf 2}, no. 3, 20 (2016).

\bibitem{Rangarajan:2018tte} 
  R.~Rangarajan,
  arXiv:1801.02648 [astro-ph.CO].

\bibitem{Li:2018wno} 
  X.~B.~Li, H.~Wang and J.~Y.~Zhu,
  Phys.\ Rev.\ D {\bf 97}, no. 6, 063516 (2018)
  [arXiv:1803.10074 [gr-qc]].

\bibitem{Herrera:2018cgi} 
  R.~Herrera,
  Eur.\ Phys.\ J.\ C {\bf 78}, no. 3, 245 (2018)
  [arXiv:1801.05138 [gr-qc]].

\bibitem{Arya:2017zlb} 
  R.~Arya, A.~Dasgupta, G.~Goswami, J.~Prasad and R.~Rangarajan,
  JCAP {\bf 1802}, no. 02, 043 (2018)
  [arXiv:1710.11109 [astro-ph.CO]].

\bibitem{Herrera:2017qux} 
  R.~Herrera,
  JCAP {\bf 1705}, no. 05, 029 (2017)
  [arXiv:1701.07934 [gr-qc]].

\bibitem{Videla:2016ypa} 
  N.~Videla,
  Eur.\ Phys.\ J.\ C {\bf 77}, no. 3, 142 (2017)
  [arXiv:1612.04124 [gr-qc]].

\bibitem{Peng:2016yvb} 
  Z.~P.~Peng, J.~N.~Yu, X.~M.~Zhang and J.~Y.~Zhu,
  Phys.\ Rev.\ D {\bf 94}, no. 10, 103531 (2016)
  [arXiv:1611.02789 [gr-qc]].


\bibitem{Goodarzi:2016iht} 
  P.~Goodarzi and H.~Mohseni Sadjadi,
  Eur.\ Phys.\ J.\ C {\bf 77}, no. 7, 463 (2017)
  [arXiv:1609.06185 [gr-qc]].

\bibitem{Levy:2016jfh} 
  A.~M.~Levy and G.~J.~Turiaci,
  Phys.\ Rev.\ D {\bf 94}, no. 8, 083514 (2016)
  [arXiv:1603.06608 [gr-qc]].

\bibitem{Visinelli:2014qla} 
  L.~Visinelli,
  JCAP {\bf 1501}, no. 01, 005 (2015)
  [arXiv:1410.1187 [astro-ph.CO]].

\bibitem{Herrera:2014mca} 
  R.~Herrera, M.~Olivares and N.~Videla,
  Int.\ J.\ Mod.\ Phys.\ D {\bf 23}, no. 10, 1450080 (2014)
  [arXiv:1404.2803 [gr-qc]].

\bibitem{Setare:2013dd} 
  M.~R.~Setare and V.~Kamali,
  JHEP {\bf 1303}, 066 (2013)
  [arXiv:1302.0493 [hep-th]].

\bibitem{Bastero-Gil:2013owa} 
  M.~Bastero-Gil, A.~Berera, T.~P.~Metcalf and J.~G.~Rosa,
  JCAP {\bf 1403}, 023 (2014)
  [arXiv:1312.2961 [hep-ph]].

\bibitem{Cerezo:2012ub} 
  R.~Cerezo and J.~G.~Rosa,
  JHEP {\bf 1301}, 024 (2013)
  [arXiv:1210.7975 [hep-ph]].
  
\bibitem{BasteroGil:2011cx} 
  M.~Bastero-Gil, A.~Berera, R.~O.~Ramos and J.~G.~Rosa,
  Phys.\ Lett.\ B {\bf 712}, 425 (2012)
  [arXiv:1110.3971 [hep-ph]].
  
\bibitem{Hall:2003zp} 
  L.~M.~Hall, I.~G.~Moss and A.~Berera,
  Phys.\ Rev.\ D {\bf 69}, 083525 (2004)
  [astro-ph/0305015].

\bibitem{Moss:2007cv} 
  I.~G.~Moss and C.~Xiong,
  JCAP {\bf 0704}, 007 (2007)
  [astro-ph/0701302].

\bibitem{Graham:2009bf} 
  C.~Graham and I.~G.~Moss,
  JCAP {\bf 0907}, 013 (2009)
  [arXiv:0905.3500 [astro-ph.CO]].

\bibitem{Ramos:2013nsa} 
  R.~O.~Ramos and L.~A.~da Silva,
  JCAP {\bf 1303}, 032 (2013)
  [arXiv:1302.3544 [astro-ph.CO]].


\bibitem{Bastero-Gil:2014jsa} 
  M.~Bastero-Gil, A.~Berera, I.~G.~Moss and R.~O.~Ramos,
  JCAP {\bf 1405}, 004 (2014)
  [arXiv:1401.1149 [astro-ph.CO]].

\bibitem{Cai:2010wt} 
  Y.~F.~Cai, J.~B.~Dent and D.~A.~Easson,
  Phys.\ Rev.\ D {\bf 83}, 101301 (2011)
  [arXiv:1011.4074 [hep-th]].

\bibitem{Bartrum:2013fia} 
  S.~Bartrum, M.~Bastero-Gil, A.~Berera, R.~Cerezo, R.~O.~Ramos and J.~G.~Rosa,
  Phys.\ Lett.\ B {\bf 732}, 116 (2014)
  [arXiv:1307.5868 [hep-ph]].

   \bibitem{BGR}
  A.~Berera, M.~Gleiser and R.~O.~Ramos,
  Phys.\ Rev.\ D {\bf 58}, 123508 (1998)
  [hep-ph/9803394].
  
  
  \bibitem{YL}
  J.~Yokoyama and A.~D.~Linde,
  Phys.\ Rev.\ D {\bf 60}, 083509 (1999)
  [hep-ph/9809409].

\bibitem{Berera:2002sp} 
  A.~Berera and R.~O.~Ramos,
  Phys.\ Lett.\ B {\bf 567}, 294 (2003)
  [hep-ph/0210301].


\bibitem{Moss:2006gt} 
  I.~G.~Moss and C.~Xiong,
  hep-ph/0603266.
  
\bibitem{BasteroGil:2010pb} 
  M.~Bastero-Gil, A.~Berera and R.~O.~Ramos,
  JCAP {\bf 1109}, 033 (2011)
  [arXiv:1008.1929 [hep-ph]].
  
\bibitem{BasteroGil:2012cm} 
  M.~Bastero-Gil, A.~Berera, R.~O.~Ramos and J.~G.~Rosa,
  JCAP {\bf 1301}, 016 (2013)
  [arXiv:1207.0445 [hep-ph]].
  
\bibitem{BasteroGil:2011mr} 
  M.~Bastero-Gil, A.~Berera and J.~G.~Rosa,
  Phys.\ Rev.\ D {\bf 84}, 103503 (2011)
  [arXiv:1103.5623 [hep-th]].


\bibitem{Bastero-Gil:2016qru} 
  M.~Bastero-Gil, A.~Berera, R.~O.~Ramos and J.~G.~Rosa,
  Phys.\ Rev.\ Lett.\  {\bf 117}, no. 15, 151301 (2016)
  [arXiv:1604.08838 [hep-ph]].

\bibitem{ArkaniHamed:2001nc} 
  N.~Arkani-Hamed, A.~G.~Cohen and H.~Georgi,
  Phys.\ Lett.\ B {\bf 513}, 232 (2001)
  [hep-ph/0105239].

\bibitem{Schmaltz:2005ky} 
  M.~Schmaltz and D.~Tucker-Smith,
  Ann.\ Rev.\ Nucl.\ Part.\ Sci.\  {\bf 55}, 229 (2005)
  [hep-ph/0502182].


\bibitem{Bastero-Gil:2017wwl} 
  M.~Bastero-Gil, S.~Bhattacharya, K.~Dutta and M.~R.~Gangopadhyay,
  JCAP {\bf 1802}, no. 02, 054 (2018)
  [arXiv:1710.10008 [astro-ph.CO]].

\bibitem{Planck}
  P.~A.~R.~Ade {\it et al.} [Planck Collaboration],
  Astron.\ Astrophys.\  {\bf 594}, A20 (2016)
  [arXiv:1502.02114 [astro-ph.CO]].

\bibitem{Bastero-Gil:2016mrl} 
  M.~Bastero-Gil, A.~Berera, R.~Brandenberger, I.~G.~Moss, R.~O.~Ramos and J.~G.~Rosa,
  JCAP {\bf 1801}, no. 01, 002 (2018)
  [arXiv:1612.04726 [astro-ph.CO]].

\bibitem{Benetti:2016jhf} 
  M.~Benetti and R.~O.~Ramos,
  Phys.\ Rev.\ D {\bf 95}, no. 2, 023517 (2017)
  [arXiv:1610.08758 [astro-ph.CO]].


\bibitem{Kapusta:2006}
J.~I.~Kapusta and C.~Gale, 
\emph{Finite-Temperature Field Theory: Principles and Applications}, 
(Cambridge University Press, Cambridge, England, 2006).

\bibitem{Cline:1997}
J.~M.~Cline and P.~A.~Lemieux, 
Phys.\ Rev.\ D {\bf 55}, 3873 (1997) 
[hep-ph/9609240].

\bibitem{Landsman}
N.~P.~Landsman and C.~van~Weert, Phys.\ Rep.\ {\bf 145}, 141 (1987); 
J.~I.~Kapusta, \emph{Finite Temperature Field Theory}, Cambridge Univ.\ Press (Cambridge 1989);   
M.~Le~Bellac, \emph{Thermal Field Theory}, Cambridge Univ.\ Press (Cambridge 1996).

\bibitem{Kamionkowski:2015yta} 
  M.~Kamionkowski and E.~D.~Kovetz,
  Ann.\ Rev.\ Astron.\ Astrophys.\  {\bf 54}, 227 (2016)
  [arXiv:1510.06042 [astro-ph.CO]].
  
\bibitem{Brandenberger:2016uzh} 
  R.~Brandenberger,
  Int.\ J.\ Mod.\ Phys.\ D {\bf 26}, no. 01, 1740002 (2016)
  [arXiv:1601.01918 [hep-th]].

\end{thebibliography}
\end{document}